\documentclass[aps,pre,reprint,superscriptaddress,longbibliography,showpacs]{revtex4-1}
\usepackage{amsmath,amssymb,graphicx,xcolor}
\usepackage[caption=false]{subfig}
\usepackage{times,bm,url}
\usepackage[colorlinks=true,breaklinks=true,allcolors=blue]{hyperref}

\newcommand{\ex}[1]{\ensuremath{\langle #1 \rangle}} %expectation value

\newcommand{\bra}[1]{\ensuremath{\langle#1\hspace{-0.04em}\vert}}
\newcommand{\ket}[1]{\ensuremath{\vert\hspace{-0.04em}#1\rangle}}

\newcommand{\mat}[1]{\begin{pmatrix}#1	\end{pmatrix}}

\DeclareMathOperator{\Tr}{{Tr}}
\DeclareMathOperator{\AdP}{{Ad_P}}
\DeclareMathOperator{\myRe}{{Re}}

\def\Oo{\ensuremath{{\cal O}}} %Order sign
\def\Uu{\ensuremath{{\cal U}}} %the unitor
\def\Dd{\ensuremath{{\cal D}}} %the dissipator 
\def\Ll{\ensuremath{{\cal L}}} %the Lindbladian
\def\Hh{\ensuremath{{\cal H}}} %the Hamiltonian
\def\Tt{\ensuremath{{\cal T}}} %iets
\def\Kk{\ensuremath{{\cal K}}} %nog iets
\newcommand{\mathtext}[1]{\quad\text{#1}\quad}

\makeatletter
\let\cat@comma@active\@empty
\makeatother

\makeatletter
\g@addto@macro\bfseries{\boldmath}
\makeatother

\usepackage{soul}

\begin{document}

\title{Bosonic representation of a Lipkin-Meshkov-Glick model with Markovian dissipation}

\author{Jan C.\ Louw}
\affiliation{Institute of Theoretical Physics,  Department of Physics, University of Stellenbosch, Stellenbosch 7600, South Africa}
\affiliation{Institute for Theoretical Physics, Georg-August Universit{\"a}t G{\"o}ttingen, 37077 G{\"o}ttingen, Germany}

\author{Michael Kastner}
\email{kastner@sun.ac.za}
\affiliation{National Institute for Theoretical Physics (NITheP), Stellenbosch 7600, South Africa}
\affiliation{Institute of Theoretical Physics,  Department of Physics, University of Stellenbosch, Stellenbosch 7600, South Africa}

\author{Johannes N.\ Kriel}
\affiliation{Institute of Theoretical Physics,  Department of Physics, University of Stellenbosch, Stellenbosch 7600, South Africa}

\date{\today}

\begin{abstract}
We study the dynamics of a Lipkin-Meshkov-Glick model in the presence of Markovian dissipation, with a focus on late-time dynamics and the approach to thermal equilibrium. Making use of a vectorized bosonic representation of the corresponding Lindblad master equation, we use degenerate perturbation theory in the weak-dissipation limit to analytically obtain the eigenvalues and eigenvectors of the Liouvillian superoperator, which in turn give access to closed-form analytical expressions for the time evolution of the density operator and observables. Our approach is valid for large systems, but takes into account leading-order finite-size corrections to the infinite-system result. As an application, we show that the dissipative Lipkin-Meshkov-Glick model equilibrates by passing through a continuum of thermal states with damped oscillations superimposed, until finally reaching an equilibrium state with a temperature that in general differs from the bath temperature. We discuss limitations of our analytic techniques by comparing to exact numerical results.
\end{abstract}

\maketitle

\section{Introduction}
Dissipation and decoherence are effects that are induced on a quantum system by its environment \cite{Hornberger09,Schlosshauer19}. These effects may be seen as a detrimental factor that is to be reduced and/or undone, and this viewpoint is taken in the majority of quantum computing schemes, as well as in other tasks in quantum information processing and storage \cite{Shor95,DevittMunroNemoto13,LidarBrun}. On the other hand, dissipation can also act as a welcome resource for quantum information tasks \cite{Kapit17}, be it for dissipation-driven quantum computation \cite{VerstraeteWolfCirac09}, for quantum error correction \cite{Reiter_etal17}, or for quantum state preparation schemes \cite{KastoryanoReiterSorensen11,Tucker_etal}. Benefits of dissipation can also be exploited in quantum many-body physics, for example to control transport in experiments with ultracold atoms \cite{Schempp_etal15,Whitlock_etal19}, or to maximize the coherence of condensate modes in the spirit of stochastic resonance \cite{WitthautTrimbornWimberger08}. All of these cases demand a thorough understanding of quantum dissipation and necessitates the development of numerical and analytical tools for its analysis. 

Compared to the already challenging task of treating uni\-tarily-evolving quantum many-body systems, the consideration of dissipative effects further adds to the technical complications. For example, exact diagonalization of $N$ unitarily evolving spin-$1/2$ degrees of freedom requires, in the absence of symmetries, to deal with matrices of size $2^N\times2^N$, whereas the dissipative case described by a Lindblad master equation requires matrices of size $2^{2N}\times2^{2N}$. In the unitary case, this scaling behavior restricts such analyses to about 14 spins on a typical (at the time of writing) personal computer, and to only 7 spins in the presence of Markovian dissipation \cite{JaschkeCarrDeVega19}, which accounts for the need to develop approximation methods that are suitable for dissipative quantum systems.

In this paper we contribute to the development of analytical approximation methods, focusing on a Lipkin-Meshkov-Glick (LMG) model subject to Markovian dissipation, as described by a Lindblad master equation \cite{BreuerPetruccione,RivasHuelga}. This model has seen recent interest in the context of experimental realizations by means of single-component Bose-Einstein condensates in a double-well potential \cite{Albiez_etal05,Levy_etal07}, or by two-component Bose-Einstein condensates in a single-well potential \cite{Zibold2010,Strobel_etal14}. While the hitherto reported experiments focus mostly on the coherent regime, dissipative effects inevitably become relevant on longer timescales. Various theoretical studies of the LMG model with Markovian dissipation have been reported, using different Lindblad master equations and focusing on different aspects and parameter regimes \cite{KhodorkovskyKurizkiVardi08,WitthautTrimbornWimberger08,Pudlik_etal13,KopylovSchaller19,Ferreira2019,LouwKrielKastner19}. 
Methods to treat these equations include exact diagonalization \cite{KopylovSchaller19}, quantum kinetic theory at the Hartree level \cite{KhodorkovskyKurizkiVardi08,WitthautTrimbornWimberger08}, and quantum jump methods \cite{Pudlik_etal13}. 

In Ref.~\cite{LouwKrielKastner19}, methods from asymptotic analysis have been used to obtain closed-form expressions for the late-time limits of expectation values of observables of interest, as well as for the rates at which those values are approached. The methods used in that paper are appealing due to their rigor and conciseness, but are also restricted in scope and applicability, especially because of the focus on the strict infinite-system limit. The present paper shares with Ref.~\cite{LouwKrielKastner19} the focus on late-time dynamics and the approach to thermal equilibrium in the LMG model with Markovian dissipation. For this scope of application, we develop a versatile approximate toolset based on a Holstein-Primakoff-type bosonic representation of spin-$S$ operators in the large-$S$ limit. This limit corresponds to a large-system limit, but unlike the techniques used in Ref.~\cite{LouwKrielKastner19}, allows us to account for leading-order finite-size corrections to the strict infinite-system limit, which has the merit of circumventing some of the pathologies of the strict infinite-$S$ system \cite{Webster2018}. The Holstein-Primakoff transformation is performed around the ground state of the infinite-system. 

The Holstein-Primakoff approximation developed in this paper permits to diagonalize the Lindblad superoperator, which in turn gives access to the time-evolution of the full density operator, and hence to expectation values of arbitrary observables, not just the selected few considered in Ref.~\cite{LouwKrielKastner19}. Our method yields well-manageable closed-form expressions. As a first application, we  show that the stationary state of the dissipative LMG model is a Gibbs thermal state, but with a temperature that in general differs from the imposed bath temperature. As a second application we calculate, for a family of initial states, the time-evolution of the density operator as well as a selection of observables of interest. Our results show that the dissipative LMG model thermalizes by passing through a continuum of thermal states on which damped oscillations are superimposed. Finally in Sec.~\ref{sec:numerics} we discuss in detail the validity as well as the limitations of our analytic techniques by comparing to exact numerical results. 

\section{Dissipative Lipkin-Meshkov-Glick model}
\label{s:model}
The physical motivation we have in mind is that of a system which, when isolated from its surroundings, evolves unitarily according to the Lipkin-Meshkov-Glick (LMG) Hamiltonian
\begin{equation}
\Hh_S=-\frac{\Lambda}{2S}S_x^2-hS_z.
\label{eq:Hs}
\end{equation}
Here $S_x$ and $S_z$ are components of a spin-$S$ vector operator, $\Lambda\geq0$ is a coupling constant, and $h$ denotes a magnetic field strength. We set $h=1$ from here on. The spin quantum number $S$ is related to the particle number $N=2S$ of the bosonic formulation in which the LMG model was originally introduced \cite{Lipkin1965}. In the semi-classical limit of large spin quantum number $S$, this model exhibits a quantum phase transition at $\Lambda=1$ from a symmetric phase to a symmetry-broken phase. In the former the model has a non-degenerate ground state with a zero $S_x$ expectation value. In the limit $N\to\infty$ the ground state in the symmetry-broken phase will be two-fold degenerate with non-zero $S_x$ expectation values \cite{Dusuel2005}.

Dissipative effects can be introduced into the model in various ways, either by {\em ad hoc}\/ procedures, or by coupling the model to an environment and tracing out the environment degrees of freedom. There is a plethora of choices of environments, of system--environment couplings, and of subsequent approximation techniques to render the resulting time-evolution equations more manageable \cite{BreuerPetruccione,RivasHuelga,KopylovSchaller19}. Here we use as an example a Lindblad master equation, derived in Ref.~\cite{LouwKrielKastner19}, for an LMG model coupled to a bosonic bath by making use of Born and Markov approximations, but avoiding the use of the secular approximation (see Appendices A and B of Ref.~\cite{LouwKrielKastner19} for details). By avoiding the secular approximation we obtain a master equation that is not {\em a priori}\/ guaranteed to have a Gibbs thermal state as its equilibrium state. This provides us with the opportunity to explore richer equilibrium properties in Sec.~\ref{sec:stationarystate} and more interesting equilibration dynamics in Sec.~\ref{sec:timeevolutionofdensitymatrix}.

We consider the master equation
\begin{equation}\label{e:Lindblad_general}
\partial_t\rho=\Ll\rho
\end{equation}
that describes the time-evolution of the density operator $\rho$.
The Lindbladian
\begin{equation}
\Ll=\Uu+\Dd
\end{equation}
consists of a unitary part
\begin{equation}\label{e:unitary}
	\Uu\rho=i\left[\rho,\Hh_S+\Hh_\gamma\right]
\end{equation}
with \footnote{Compared to $\Hh_\gamma$ in \cite{LouwKrielKastner19} we have neglected some terms which only contribute to $\Ll$ at order $\Oo(S^{-1})$.}
\begin{equation}\label{e:Hgamma}
	\Hh_\gamma=\frac{\gamma}{4S}\left\{S_x,S_y\right\},
\end{equation}
and a dissipative part
\begin{equation}\label{e:D}
	\Dd\rho=L\rho L^\dag-\frac{1}{2}\left\{L^\dag L,\rho\right\}
\end{equation}
with jump operator
\begin{equation}\label{e:jump}
	L=\sqrt{\frac{2\gamma T}{S}}\left(S_x+\frac{i}{4T}S_y\right),
\end{equation}
where $T$ denotes temperature of the environment. Combining the above, the Lindblad equation reads
\begin{equation}
	\partial_t\rho=\Ll\rho=i\left[\rho,\Hh_S+\Hh_\gamma\right]+L\rho L^\dag-\frac{1}{2}\left\{L^\dag L,\rho\right\}.
	\label{eq:lindblad1}
\end{equation}
When deriving Eqs.~\eqref{e:Lindblad_general}--\eqref{eq:lindblad1} from a microscopic model of system and environment, the nonnegative parameter $\gamma$ in Eqs.~\eqref{e:Hgamma} and \eqref{e:jump} is a measure of the system--environment coupling strength and is assumed to be small. Since in the present work we are not concerned with such a microscopic point of view, we ``postulate'' Eqs.~\eqref{e:Lindblad_general}--\eqref{eq:lindblad1} as our dissipative LMG model and take the liberty to admit arbitrary nonnegative values of $\gamma$. While we work with these specific equations for concreteness, the bosonization and vectorization techniques developed in Sec.~\ref{s:bosonization} can be applied to a broad range of Lindblad master equations describing large-$S$ spin models with dissipation.

\section{Bosonization of the Lindblad Equation}
\label{s:bosonization}
Our treatment of the Lindblad equation \eqref{eq:lindblad1} uses of a combination of bosonization and vectorization procedures, leading eventually to a representation of the Lindbladian $\Ll$ as an operator acting on a two-mode bosonic Fock space, the elements of which represent density operators. This form of $\Ll$ is amenable to standard techniques, which provide insight into the system's stationary state and its equilibration dynamics.  

\subsection{Holstein-Primakoff mapping}
Our first step is common to many treatments of the LMG model. Using the Holstein-Primakoff (HP) mapping \cite{Holstein1940} we introduce representations of the spin operators appearing in $\Hh_S$ in terms of bosonic creation and annihilation operators. These operators describe the low-lying excitations above the system's semi-classical ground state. Accordingly, the properties of this ground state, most notably its magnetization, steer the construction of the HP mapping.

In the large-$S$ limit a simple variational calculation \cite{RibeiroVidalMosseri08} produces the magnetization 
\begin{equation}\label{eq:magnetization}
	\bm{m}=\ex{\bm{S}}/S=(\sin\theta_0,0,\cos\theta_0)
\end{equation}
with 
\begin{equation}\label{e:theta0}
\theta_0=\begin{cases} 0 & \text{for $0\leq\Lambda<1$},\\
\pm\arccos(1/\Lambda) & \text{for $1\leq\Lambda$},
\end{cases}
\end{equation}
where the expectation value is taken with respect to the semi-classical ground state. The first of the two cases in \eqref{e:theta0} corresponds to the symmetric phase with a vanishing $x$-component of the magnetization. In the second case, the two possible signs for $\theta_0$, and hence for $m_x=\sin\theta_0$, reflect the doubly degenerate ground state of the symmetry-broken phase. Only one of these two states can be selected as reference point for the HP construction, which introduces an explicit breaking of the $P=\exp(i\pi S_z)$ parity symmetry present in $\Hh_S$. As a result, the description of the dynamics derived from this mapping is unable to capture effects due to tunneling between the two ground states. A more detailed discussion of this point will follow in Secs.~\ref{sec:symmetries} and \ref{sec:numerics}.

The HP mapping is applied to a spin operator $\bm{S}'=(S_x',S_y',S_z')$ that corresponds to an axis system in which the $z'$-direction is aligned with the ground state magnetization $\bm{m}$. It follows that $\bm{S}$ and $\bm{S}'$ are related by $\bm{S}=R_y\bm{S}'$, where
\begin{equation}
	R_y=\mat{\cos\theta_0 & 0 & \sin\theta_0\,\\ 0 & 1 & 0 \\ -\sin\theta_0 & 0 &\cos\theta_0}
\end{equation}
is a rotation around the $y$ axis. As desired, the magnetization in the rotated frame is $\bm{m}'=\ex{\bm{S}'}/S=(0,0,1)$. Since we have $\theta_0=0$ in the symmetric phase, the rotation $R_y$ is non-trivial only in the symmetry-broken phase. Also, the rotation relating $\bm{S}$ and $\bm{S}'$ is dictated by $\Hh_S$ alone, and it is not obvious that the presence of $\Hh_\gamma$ in $\Uu$, or indeed the dissipator $\Dd$ itself, should not play a role here. It will be shown below that terms in $\Hh_\gamma$ that would necessitate a different choice of $\bm{S}'$ are canceled exactly by terms originating from the dissipator.

The HP mapping amounts to expressing the components of $\bm{S}'$ in terms of the bosonic creation and annihilation operators $a$ and $a^\dag$ according to \cite{Holstein1940}
\begin{equation}
		S_+'=\sqrt{2S-a^\dag a}a \mathtext{and} S_z'=S-a^\dag a.
		\label{eq:HPMapping}
\end{equation}
It is straightforward to verify that, for $S_+'$, $S_-'=(S_+')^\dag$, and $S_z'$ satisfying spin commutation relations, this definition guarantees that $a$ and $a^\dag$ satisfy bosonic commutation relations. Combining \eqref{eq:HPMapping} with the relation $\bm{S}=R_y\bm{S}'$ allows the constituents of the Lindbladian, namely $\Hh_S$, $\Hh_\gamma$ and $L$, to be expressed in terms of bosonic operators. In the resulting expressions we then expand the square roots from $S_\pm'$ in orders of $1/S$, keeping only terms which may contribute to $\Ll$ at orders $\mathcal{O}(S)$, $\mathcal{O}(S^{1/2})$, and $\mathcal{O}(S^0)$. For $\Hh_S$ \eqref{eq:Hs} this yields
\begin{equation}\label{eq:Hsa}
	\Hh_S=\omega_a a^\dag a+\Gamma_a\left(a a+a^\dag a^\dag\right)+\delta_0
\end{equation}
where
\begin{subequations}
\begin{align}
	\omega_a&=\left(m_z+\Lambda-3m_z^2\Lambda/2\right),\\
	\Gamma_a&=-m_z^2\Lambda/4,\\
	\delta_0&=-S(m_z+\Lambda m_x^2/2)-\Lambda m_z^2/4,
\end{align}
\end{subequations}
and $m_x$ and $m_z$ are the semi-classical (infinite-$S$) values given in Eqs.~\eqref{eq:magnetization} and \eqref{e:theta0}. Applying the same treatment to $\Hh_\gamma$ \eqref{e:Hgamma} produces
\begin{equation}\label{e:Hgammabosonic}
	\Hh_\gamma=\frac{im_x\gamma\sqrt{2S}}{4}(a^\dag-a)+\frac{im_z\gamma}{4}\left(a^\dag a^\dag-a a\right),
\end{equation}
where we have dropped scalar terms, as they do not contribute to $\Ll$. For the jump operator \eqref{e:jump} we find
\begin{align}\label{e:Lbosonic}
	L=&m_x\sqrt{2\gamma TS}+\frac{1}{4}\sqrt{\frac{\gamma}{T}}\left[(4m_z T-1)a^\dag+(4m_z T+1)a\right]\nonumber\\
	&-m_x\sqrt{\frac{2\gamma T}{S}}a^\dag a.
\end{align}
The first terms in \eqref{e:Hgammabosonic} and \eqref{e:Lbosonic} are of order $\mathcal{O}(S^{1/2})$ and, at least individually, might contribute nontrivially to $\Ll\rho$. This may appear to rule out taking the large-$S$ limit on the level of the Lindblad equation. We show next that, on closer inspection, the contributions from these terms to $\Ll$ actually cancel. To see this, first note that there is some freedom in how the unitary term and the dissipator in $\Ll$ are identified. In fact, for any scalar $c$ we can write $\Ll$ from Eq.~\eqref{eq:lindblad1} in the form 
\begin{align}
	\Ll\rho=i[\rho,\Hh_S+\Hh'_\gamma]+L'\rho L'^\dag-\tfrac{1}{2}\{L'^\dag L',\rho\},
\end{align}
where $L'=L-c$, $\Hh'_\gamma=\Hh_\gamma+\Hh_c$, and
\begin{equation}
	\Hh_c=\tfrac{i}{2}\bigl(c^* L-c L^\dag\bigr).
\end{equation}
This is due to a cancellation of $c$-dependent terms between the modified unitary term and dissipator. This allows us to shift the jump operator by a scalar, and compensate for this by adding $\Hh_c$ to the generator of the unitary dynamics. If we chose $c=m_x\sqrt{2\gamma TS}$ then this would eliminate the problematic term from $L$. For this choice of $c$ we find that $\Hh_c=-im_x\gamma\sqrt{2S}(a^\dag-a)/4$, which in turn cancels the $\mathcal{O}(S^{1/2})$ term in $\Hh'_\gamma$. Furthermore, with the $\mathcal{O}(S^{1/2})$ term absent from $L'$, the final $\mathcal{O}(S^{-1/2})$ term can only contribute at this same order to the Lindbladian, and may therefore be dropped. Combining these results, we conclude that we may proceed with the original Lindblad equation \eqref{eq:lindblad1}, using \eqref{eq:Hsa},
\begin{equation}
	\Hh_\gamma=\frac{im_z\gamma}{4}\left(a^\dag a^\dag-a a\right),
	\label{eq:finalHgamma}
\end{equation}
and
\begin{equation}
	L=\frac{1}{4}\sqrt{\frac{\gamma}{T}}\left[(4m_z T-1)a^\dag+(4m_z T+1)a\right].
	\label{eq:finalL}
\end{equation}

\subsection{Diagonalising \texorpdfstring{$\Hh_S$}{HS}}
Before dealing with the Lindbladian it will be useful to perform a Bogoliubov transformation to bring $\Hh_S$ into diagonal form. This introduces a new species of $b$-bosons defined by
\begin{equation}
	a=\sinh(\phi_b/2)b^\dag+\cosh(\phi_b/2)b,
	\label{eq:bogoliubov}
\end{equation}
where the Bogoliubov angle $\phi_b$ is set according to 
\begin{equation}
	\tanh\phi_b=\epsilon\mathtext{with}\epsilon=-2\Gamma_a/\omega_a.
	\label{eq:bogoliubovangle1}
\end{equation}
Substituting \eqref{eq:bogoliubov} and its adjoint into $\Hh_S$ produces the diagonal Hamiltonian
\begin{equation}
	\Hh_S=\omega_b b^\dag b+E_{0},
	\label{eq:HSdiagonal}
\end{equation}
where $E_{0}=\delta_0+(\omega_b-\omega_a)/2$ and $\omega_b=\omega_a\sqrt{1-\epsilon^2}$. The latter parameter can be simplified to
\begin{equation}
	\omega_b=\begin{cases} \sqrt{1-\Lambda} & \text{for $0\leq\Lambda<1$},\\
\sqrt{\Lambda^2-1} & \text{for $1\leq\Lambda$},
\end{cases}
\end{equation}
where the expressions \eqref{eq:magnetization} and \eqref{e:theta0} for the semi-classical magnetization components have been used. To rewrite the Lindblad equation \eqref{eq:lindblad1} entirely in terms of the Bogoliubov $b$-bosons, we apply the transformation \eqref{eq:bogoliubov} and \eqref{eq:bogoliubovangle1} to $\Hh_\gamma$ in \eqref{eq:finalHgamma}, which yields
\begin{equation}
	\Hh_\gamma=\frac{im_z\gamma}{4}\left(b^\dag b^\dag-b b\right),
\end{equation}
and to $L$ in \eqref{eq:finalL}, which gives
\begin{equation}
	L=\sqrt{\gamma}\left(B_+ b^\dag+B_- b\right)
	\label{eq:newL}
\end{equation}
with
\begin{multline}
	B_\pm=\frac{1}{4\sqrt{T}}[(4m_zT\pm1)\sinh(\phi_b/2)\\
	+(4m_zT\mp1)\cosh(\phi_b/2)].
\end{multline}
The expression for $B_\pm$ can be simplified by inserting the solution for $\phi_b$ from \eqref{eq:bogoliubovangle1} and considering the two phases individually. In both phases $B_\pm$ is found to reduce to 
\begin{equation}
	B_\pm=\sqrt{m_z}\left(\sqrt{\frac{T}{\omega_b}}\mp\frac{1}{4}\sqrt{\frac{\omega_b}{T}}\right).
	\label{eq:bplusminus}
\end{equation}
For later use we note that
\begin{equation}\label{eq:BIdentity}
	B_-^2-B_+^2=m_z
\end{equation}
and
\begin{equation}
	\frac{B_+}{B_-}=\frac{4T-\omega_b}{4T+\omega_b}.
	\label{eq:BIdentity2}
\end{equation}

\subsection{Vectorizing the Lindblad equation}
\label{sec:vectorizing}
In its present form the Lindbladian $\mathcal{L}$ is a superoperator, acting on the density operator $\rho$, which in turn acts on the single mode bosonic Fock space $\mathcal{B}_1$. There exists a natural mapping between operators acting on $\mathcal{B}_1$ and elements of the two-mode bosonic Fock space $\mathcal{B}_2=\mathcal{B}_1\otimes\mathcal{B}_1$. This allows us to represent the density operator $\rho$ as a vector $\ket{\rho}$ in $\mathcal{B}_2$, while the Lindbladian becomes an operator acting on this space. Working in the basis of $b$-boson number states, this mapping amounts to
\begin{equation}
	\rho=\sum_{ij}\rho_{ij}\ket{i}\bra{j}\longleftrightarrow\ket{\rho}=\sum_{ij}\rho_{ij}\ket{i}\otimes\ket{j}.
\end{equation}
Under this map the left and right action of operators $A=A(b,b^\dag)$ and $B=B(b,b^\dag)$ on $\rho$ become
\begin{equation}
	A\rho B\longleftrightarrow A\otimes B^T\ket{\rho},
\end{equation}
where the transpose operation ($T$) exchanges $b$ and $b^\dag$, but leaves scalars unaffected. The unitary term
\begin{equation}
	\Uu\rho=i[\rho,\Hh_S+\Hh_\gamma]
\end{equation}
represented as an operator on $\mathcal{B}_2$ reads
\begin{equation}
	\Uu=i\mathbb{I}\otimes(\Hh_S+\Hh_\gamma)^T-i(\Hh_S+\Hh_\gamma)\otimes\mathbb{I},
\end{equation}
while the dissipator
\begin{equation}
	\Dd\rho=L\rho L^\dag-\tfrac{1}{2}\bigl\{L^\dag L,\rho\bigr\}
\end{equation}
becomes
\begin{equation}
\Dd=L \otimes L^* - \tfrac{1}{2} \bigl(L^\dag L \otimes \mathbb{I} + \mathbb{I}\otimes L^T L^*\bigr).
\end{equation}
Here the conjugation operation ($*$) only affects scalars and leaves the boson operators unchanged. We see from \eqref{eq:newL} and \eqref{eq:bplusminus} that $L^*=L$, and so $L^T=L^\dag$. Defining
\begin{align}
b_{1}&=b\otimes\mathbb{I},& b_{2}&=\mathbb{I}\otimes b,
\end{align}
the final forms of $\Uu$ and $\Dd$ read
\begin{equation}
	\Uu=i\omega_b(b_2^\dag b_2-b_1^\dag b_1)+\frac{m_z\gamma}{4}\left(b_1^\dag b_1^\dag+b_2^\dag b_2^\dag-\text{h.c.}\right)
	\label{eq:bosonisedU}
\end{equation}
and
\begin{equation}
\Dd=L_1L_2-\tfrac{1}{2}\bigl(L_1^\dag L_1 + L_2^\dag L_2\bigr),
\label{eq:bosonisedD}
\end{equation}
where 
\begin{equation}\label{eq:bosonisedL}
	L_i=\sqrt{\gamma}\bigl(B_+b_i^\dag+B_-b_i\bigr).
\end{equation}

\subsection{Small-\texorpdfstring{$\gamma$}{gamma} perturbative diagonalization of the Lindbladian}
\label{sec:lindbladdiagonalise}
We have arrived at a representation of the Lindbladian $\Ll=\Uu+\Dd$ as an operator acting on a two-mode bosonic Fock space. This operator is quadratic in creation and annihilation operators, which suggests to attempt an exact diagonalization via a Bogoliubov transformation. The approach presented in \cite{prosen2010} provides a systematic way of constructing the new species of bosons required for this task. However, within the weak coupling regime a simpler, perturbative approach will suffice, one which also allows us to exploit the simple algebraic properties of the operators appearing in $\Ll$. Section \ref{sec:strongcoupling} contains a brief discussion of the results that the approach of \cite{prosen2010} produces in the strong coupling regime.

We proceed on the basis of standard perturbation theory and write $\Ll=\Ll_0+\gamma\Ll'$, where 
\begin{equation}
	\Ll_0=i\omega_b(b_2^\dag b_2-b_1^\dag b_1),
	\label{eq:L0Def}
\end{equation}
while
\begin{multline}
	\gamma\Ll'=\frac{m_z\gamma}{4}\left(b_1^\dag b_1^\dag+b_2^\dag b_2^\dag-\text{h.c.}\right)\\+L_1L_2-\frac{1}{2}\left(L_1^\dag L_1 + L_2^\dag L_2\right)
	\label{eq:LPrimeDef}
\end{multline}
represents the perturbation. The spectrum of $\Ll_0$ is highly degenerate, and the eigenspace corresponding to a certain eigenvalue is spanned by $b_1$- and $b_2$-boson Fock states with a fixed boson number difference. We must therefore diagonalize $\Ll'$ within each of these subspaces individually. Let $\Delta$ represent the value of $b_2^\dag b_2-b_1^\dag b_1$, and consider the projection $\Ll'_\Delta$ of $\Ll'$ into this subspace. Performing this projection amounts to dropping terms which do not conserve $\Ll_0$. This yields
\begin{align}\label{e:LPrimeDelta}
\Ll'_\Delta=&m_z/2-(B_+^2+B_-^2)\tfrac{1}{2}\left(b_1^\dag b_1+b_2^\dag b_2+1\right)\nonumber
\\&+B_+^2b_1^\dag b_2^\dag+B_-^2 b_1 b_2,
\end{align}
where we have used \eqref{eq:BIdentity} to obtain the $m_z/2$ term. The structure of $\Ll'_\Delta$ is reminiscent of a pairing Hamiltonian, albeit a non-Hermitian one. In Appendix \ref{sec:su11diagonalisation} we detail the construction of a similarity transformation $\Tt$ which brings this operator into diagonal form. This construction is aided by the fact that the operators appearing in $\Ll'_\Delta$ obey ${\rm su}(1,1)$ commutation relations, and therefore transform in a simple way under ${\rm SU}(1,1)$ group transformations. While this construction is equivalent to performing a non-unitary Bogoliubov transformation, having an explicit form for $\Tt$, given in Eqs. \eqref{eq:T1def} and \eqref{eq:T2def}, also provides us with direct access to the eigenstates. Applying $\Tt$ produces
\begin{equation}
	\Tt^{-1}\Ll'_\Delta \Tt=-\frac{m_z}{2}\left(b_1^\dag b_1+b_2^\dag b_2\right).
\end{equation}
For a fixed $\Delta$ the eigenvalues of $b_1^\dag b_1+b_2^\dag b_2$ are $|\Delta|+2n$ for $n\in\mathbb{N}_0$, and the corresponding eigenstates are combined $b_1$- and $b_2$-boson Fock states. It is straightforward to check that the unperturbed part \eqref{eq:L0Def} is invariant under the action of $\Tt$. Combining these results, we conclude that in the weak coupling regime the eigenvalues of $\Ll$ are
\begin{equation}
	\lambda_{\Delta,n}=i\omega_b\Delta-\frac{m_z\gamma}{2}\left(|\Delta|+2n\right)
	\label{eq:Leigenvalues}
\end{equation}
with $\Delta\in\mathbb{Z}$ and $n\in\mathbb{N}_0$. The eigenstates are the transformed Fock states
\begin{equation}
	\ket{\rho_{\Delta,n}}=\begin{cases} \Tt\ket{n,n+\Delta} & \text{for $\Delta\geq0$},\\
\Tt\ket{n-\Delta,n} & \text{for $\Delta<0$}.
\end{cases}
	\label{eq:Leigenstates}
\end{equation}
Note that the bath temperature $T$ does not feature in the eigenvalues $\lambda_{\Delta,n}$, but enters into the eigenstates via the transformation $\Tt$.

\subsection{Stationary state}
\label{sec:stationarystate}

An eigenstate of the Lindblad superoperator $\Ll$ with zero eigenvalue is a stationary state of the Lindblad equation \eqref{eq:lindblad1}. In the vectorized language of Secs.~\ref{sec:vectorizing} and \ref{sec:lindbladdiagonalise}, it therefore follows that $\ket{\rho_{0,0}}=\Tt\ket{0,0}$ is the unique stationary state. (In the symmetry-broken phase this statement should be qualified further; see Secs.~\ref{sec:symmetries} and \ref{sec:numerics}.) Since the nonzero eigenvalues of $\Ll$ all have negative real parts, the system always equilibrates to $\ket{\rho_{0,0}}$. At long times the equilibration rate is set by $|\myRe({\lambda_{\pm1,0}})|=m_z\gamma/2$, i.e., by the slowest decay rate of the nonstationary eigenstates of $\Ll$. 

In order to assess whether the stationary state is a Gibbs thermal state, it is convenient to convert the vectorized state $\ket{\rho_{0,0}}$ into operator form. In Appendix\!\! \ref{sec:su11factorisation} it is shown that $\ket{\rho_{0,0}}=\Tt\ket{0,0}$ can be simplified to produce, up to normalization,
\begin{equation}
	\ket{\rho_{0,0}}\propto e^{(B_+/B_-)^2 b_1^\dag b_2^\dag}\ket{0,0}=\sum_{n=0}^\infty \left(\frac{B_+}{B_-}\right)^{2n}\ket{n,n}.
	\label{eq:rho00vector}
\end{equation}
Replacing $\ket{n,n}$ by $\ket{n}\bra{n}$ gives the operator form of the stationary state,
\begin{equation}\label{eq:rho00}
	\rho_{0,0}\propto \exp\left[2\ln(B_+/B_-)b^\dag b\right]\propto \exp\left(-\Hh_S/T_\text{ss}\right),
\end{equation}
where, up to a scalar term, we have identified $\Hh_S$ with its diagonal form in \eqref{eq:HSdiagonal}, and defined the temperature
\begin{equation}\label{e:Tss}
	T_\text{ss}=-\frac{\omega_b}{2\ln(B_+/B_-)}.
\end{equation}
Comparing $\rho_{0,0}$ to the Gibbs state $\rho_{\rm th}=\exp(-\Hh_S/T)$ now amounts to a comparison of $T_\text{ss}$ with the bath temperature $T$.  Using the expression for $B_+/B_-$ in Eq. \eqref{eq:BIdentity2} we obtain the expansion 
\begin{equation}\label{e:Tss_expansion}
\frac{T_\text{ss}}{T}=1-\frac{\omega_b^2}{48T^2}-\frac{\omega_b^4}{2880T^4}+\mathcal{O}\left(\frac{\omega_b^6}{T^6}\right).
\end{equation}
Already for $T\gtrsim 2\omega_b$ it is clear that $T_\text{ss}$ will match the bath temperature $T$ very closely, and so within this regime the stationary state $\rho_{0,0}$ in \eqref{eq:rho00} indeed coincides with the thermal Gibbs state.

\subsection{Time-evolution of the density operator}
\label{sec:timeevolutionofdensitymatrix}

The perturbative results of Sec.~\ref{sec:lindbladdiagonalise} can be used to calculate, for a given initial $\rho(0)$, the time-evolution $\rho(t)$ of the density operator, which in turn gives access to the time-evolution of expectation values of arbitrary observables. To this aim, we recall that the perturbative calculation amounted to diagonalizing the operator
\begin{equation}
\Ll_\Delta\equiv \Ll_0+\gamma\Ll'_\Delta,
\end{equation}
with $\Ll_0$ and $\Ll'_\Delta$ as defined in Eqs.~\eqref{eq:L0Def} and \eqref{e:LPrimeDelta}. The algebraic properties of the operators appearing in $\Ll_\Delta$ allow us to apply $\exp[t\Ll_\Delta]$ directly to certain types of initial states.

As an illustration, we consider the evolution of the initial state
\begin{equation}\label{e:rho0prime}
\rho(0)=\ket{\psi}\bra{\psi},
\end{equation}
where
\begin{equation}\label{eq:psi}
\ket{\psi}=R_y(\theta)\ket{\text{GS}}
\end{equation}
is the ground state of the system Hamiltonian $\Hh_S$, rotated using
\begin{equation}\label{e:Ry}
R_y(\theta)=\exp{[-i\theta S_y]}
\end{equation}
by an angle $\theta$ around the $y$-axis. For simplicity we consider the symmetric phase in which $m_z=1$. It is now possible to apply $\exp[t\Ll_\Delta]$ to the vectorized initial state $\ket{\rho(0)}$ and to bring the result into a simple form. The details of this calculation are shown in Appendix \!\ref{sec:su11evolution}. After switching back to the nonvectorized language the time-evolved density operator is given by
\begin{equation}
 \rho(t)=U(t)\rho_\text{th}(t)U^\dag(t)
 \label{eq:evolvedrho}
\end{equation}
with
\begin{subequations}
\begin{align}
	U(t)&=\exp\bigl[-i\theta e^{-\gamma t/2}\left(\cos(\omega_b t) S_y+e^{-\theta}\sin(\omega_b t)S_x\right)\bigr],\\
	\rho_\text{th}(t)&=\bigl(1-e^{-\omega_b/T_S(t)}\bigr)e^{-\Hh_S/T_S(t)},\\
	\frac{1}{T_S(t)}&=\frac{1}{\omega_b}\log\left(\frac{e^{-\omega_b/T_\text{ss}}-e^{-\gamma t}}{1-e^{-\gamma t}}\right).
	 \label{eq:TSdefinition}
\end{align}
\end{subequations}
The form of $\rho(t)$ in \eqref{eq:evolvedrho} provides a simple and intuitive picture of the dynamics which leads the system to thermal equilibrium. The density matrix $\rho_\text{th}(t)$ represents a thermal state with a time-dependent temperature $T_S(t)$. The latter increases from $T_S(0)=0$, approaching a final value of $T_\text{ss}$, the steady state temperature identified in Eq.~\eqref{e:Tss}. As was shown in Eq.~\eqref{e:Tss_expansion}, $T_\text{ss}$ is essentially equal to the bath temperature $T$ when $T\gtrsim 2\omega_b$. The behavior of $T_S(t)$ reflects the heating of the system by the bath, and is independent of the rotation angle $\theta$ that characterizes the initial state. In parallel with this heating process, the unitary transformation $U(t)$ introduces an oscillating and damped rotation into $\rho(t)$. These oscillations result from the $R_y(\theta)$ rotation in $\rho(0)$, which introduces a misalignment between the initial state's magnetization and that of the stationary state.

\begin{figure}
	\centering
	\includegraphics[width=1\linewidth]{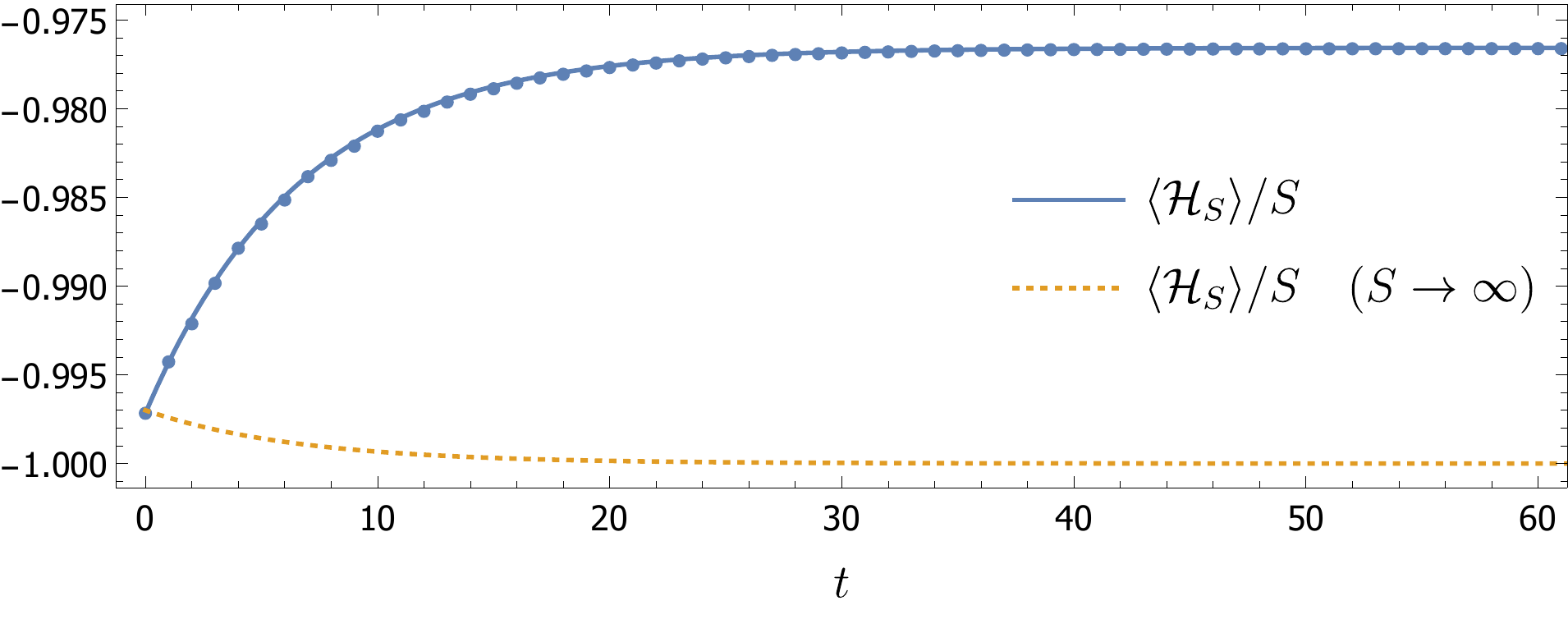}
	\includegraphics[width=1\linewidth]{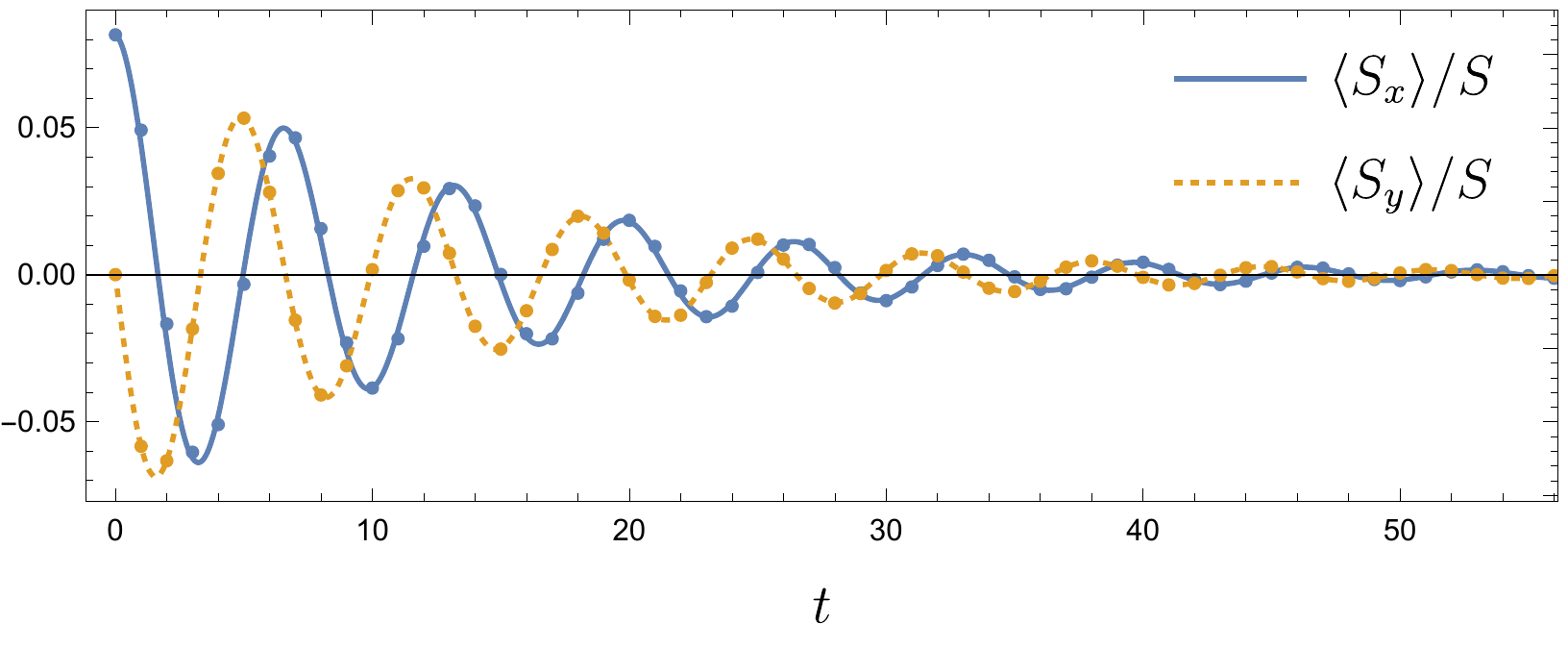}
	\caption{\label{fig:energydynamicsplot}
	Results for the expectation value of the system Hamiltonian $\Hh_S$ and for the $m_x$ and $m_y$ components of the magnetization, plotted as functions of time for the initial state described in the text. Solid and dashed lines correspond to the predictions of Eqs.~\eqref{eq:HsEVAnalytic}, \eqref{eq:mxEVAnalytic} and \eqref{eq:myEVAnalytic}. In the top figure the prediction of Eq.~\eqref{eq:HsEVAnalytic} is shown both with (solid line) and without (dashed line) the finite-size correction terms. The dots show exact numerical results obtained by solving the original spin-based Lindblad equation \eqref{eq:lindblad1} for $S=150$. Parameters were set to $\Lambda=0.1$, $\gamma=0.15$, $T=4$, and $\theta=1/\sqrt{S}$.}
\end{figure}

From the explicit form of $\rho(t)$ in \eqref{eq:evolvedrho}, various quantities of interest can be calculated. For the system energy we obtain the expectation value
\begin{equation}
	\frac{\ex{\Hh_S}}{S}=\frac{E_{0}}{S}+\frac{\theta^2 \omega_b}{2}e^{-\gamma t-\phi_b}+\frac{\omega_b}{S(e^{\omega_b/T_{S}(t)}-1)},
	\label{eq:HsEVAnalytic}
\end{equation}
which provides a nice illustration of the two processes described above: The second term on the right-hand side of \eqref{eq:HsEVAnalytic} describes the dissipation of the energy imparted to the system by the $R_y(\theta)$ rotation in the initial state. The last term on the right-hand side of \eqref{eq:HsEVAnalytic} is the heat absorbed from the bath, and represents a temperature-dependent finite-size correction to the $S\rightarrow\infty$ limit of $\ex{\Hh_S}/S$. In \cite{LouwKrielKastner19} the thermalization of this system was studied using a set of semi-classical equations of motion for the spin components. This approach provided a description of the dynamics far away from equilibrium, unlike the present local description which follows from the Holstein-Primakoff mapping. However, the results of \cite{LouwKrielKastner19}, being derived in the strict large-$S$ limit, do not account for quantum fluctuations, nor any thermal effects. In fact, the predictions of \cite{LouwKrielKastner19} coincide with the $S\rightarrow\infty$ limit of the expression in \eqref{eq:HsEVAnalytic}. Figure \ref{fig:energydynamicsplot} shows of the prediction of the present approach with that of \cite{LouwKrielKastner19}, together with exact numerical results for $S=150$. The inclusion of the finite-size corrections clearly lead to much better agreement with the exact results, which is one of the main merits of the bosonization approach advocated in the present work.

For the components of the magnetization
\begin{equation}
\bm{m}(t)=\Tr(\bm{S}\rho(t))/S
\end{equation}
we find
\begin{subequations}
\begin{align}
	m_x(t)=&\theta e^{-\gamma t/2}\cos(\omega_b t)\label{eq:mxEVAnalytic}\\
	m_y(t)=&-\theta e^{-\gamma t/2-\phi_b}\sin(\omega_b t)\label{eq:myEVAnalytic}\\
	m_z(t)=&1-\frac{\theta^2}{2}e^{-\gamma t-\phi_b}\left[\cosh\phi_b+\cos(2\omega_b t)\sinh\phi_b\right]\nonumber\\
	&-\frac{1}{S}\left(\sinh^2(\phi_b/2)+\frac{\cosh\phi_b}{e^{\omega/T_S(t)}-1}\right).
\end{align}
\end{subequations}
As expected, the rotation operator entering in \eqref{eq:evolvedrho} generates damped oscillations in the three spin components. Time-dependent expectation values of the other observables (besides the above treated magnetization components) can be derived along similar lines.

\subsection{Diagonalization for arbitrary dissipation strength \texorpdfstring{$\gamma$}{gamma}}
\label{sec:strongcoupling}
The perturbative approach of Sec.~\ref{sec:lindbladdiagonalise} relies on the requirement that the system--bath coupling is weak. However, even when the weak-coupling condition $\gamma\ll\omega_b$ is violated, an analytic treatment of the bosonic Lindblad equation is still possible by the method of third quantization \cite{prosen2010}. This approach allows for the exact diagonalization of bosonic Lindbladians in which the Hamiltonian and jump operators are respectively quadratic and linear in the boson creation and annihilation operators. When applied to the bosonised Lindblad operator $\Ll=\Uu+\Dd$ given by \eqref{eq:bosonisedU}--\eqref{eq:bosonisedL}, the primary outputs of the third quantization procedure are encoded in the so-called rapidities
\begin{equation}
	\beta_\pm=\frac{1}{4}\left(m_z\gamma\pm i\sqrt{4\omega_b^2-m_z^2\gamma^2}\right)
	\label{eq:rapidities}
\end{equation}
and a complex symmetric matrix $Z$ with elements
\begin{subequations}
\begin{align}
	Z_{11}&=Z_{22}=\frac{m_z\gamma(m_z\gamma-2\omega_b i)}{32 T \omega_b},\\
	Z_{12}&=\frac{m_z^2\gamma^2+2(4T-\omega_b)^2}{32 T \omega_b}.
\end{align}
\end{subequations}
The physical content of these quantities is as follows. The eigenvalues of the Lindblad operator are given in terms of the rapidities by $-2(\beta_+n_++\beta_-n_-)$ where $n_\pm\in\mathbb{Z}$. A Taylor expansion in $\gamma$ confirms that this result is consistent with the eigenvalues found in \eqref{eq:Leigenvalues} in the weak-coupling limit. The entries of $Z$ determine the stationary state expectation values,
\begin{align}
	\ex{b^\dag b}&=Z_{12},& \ex{bb}&=Z_{11},
	\label{eq:prosenexpval}
\end{align}
and other expectation values follow via Wick's theorem. Setting $\gamma=0$ in these expressions recovers the thermal state \eqref{eq:rho00}. Increasing $\gamma$ results in deviations from these thermal values. This does not come as a surprise, as a large value of $\gamma$ invalidates the Born-Markov approximation upon which the derivation of the Lindblad equation is based \cite{LouwKrielKastner19}, and hence severs the connection to the microscopic model. However, the stationary state does not exhibit any nonanalytic behavior or instabilities with increasing $\gamma$, which appears surprising in light of the fact that the generator $\Hh_S+\Hh_\gamma$ of the unitary evolution becomes an unstable inverted oscillator when $m_z\gamma>2\omega_b$. The spectrum of $\Ll$ undergoes an interesting qualitative change at $m_z\gamma>2\omega_b$, where all eigenvalues become real, as can be seen from Eq.~\eqref{eq:rapidities}. This eliminates the oscillatory behavior from the dynamics generated by $\Ll$, leading to an overdamped decay. The simple dependence of the Lindblad dynamics on $\gamma$, devoid of nonanalyticities and instabilities, can be understood as a consequence of cancellations of $\gamma$-dependent terms in the unitary part $\Hh_\gamma$ of $\Ll$ and in the dissipator. 

It would be interesting to consider a class of Lindblad equations for the LMG model in which the coupling enters only in the dissipator. In this case it is conceivable that strong coupling modifies the model's phase structure and impacts on the stability of the stationary state, similar to the findings of Ref.~\cite{SinhaSinha19} in a closed-system setting modeled by quadratic system and bath Hamiltonians.

\section{Parity symmetry of \texorpdfstring{$\Hh_S$}{HS} and \texorpdfstring{$\Ll$}{L}}
\label{sec:symmetries}

Parity symmetry breaking is essential for the understanding of the phase transition occurring in the LMG model in the thermodynamic limit for $\Lambda>1$. In this section we discuss the role of parity symmetry in the Hamiltonian as well as the Lind\-blad\-ian. The breaking of this symmetry strongly affects the equilibration dynamics of the LMG model, and a thorough understanding is helpful for clarifying the status of the analytic results in Secs.~\ref{sec:lindbladdiagonalise} and \ref{sec:stationarystate}. We consider integer values of $S$ for simplicity.

The LMG Hamiltonian \eqref{eq:Hs} commutes with the parity operator
\begin{equation}
P=\exp(i\pi S_z),
\end{equation}
and therefore the eigenstates of $\Hh_S$ can be chosen to have well-defined parity. Since $P^\dag S_x P=-S_x$, such states necessarily have zero $S_x$ expectation values. 

The spectrum of the LMG Hamiltonian has previously been analyzed using a variety of approaches. However, analytic results are typically restricted to the large-$S$ limit \cite{RibeiroVidalMosseri07,RibeiroVidalMosseri08} or low energies \cite{Dusuel2005}, and many studies also consider only one of the two parity sectors. Our interest lies with large but finite $S$, and in how the spectra of the odd and even parity sectors compare. To this end, numerical results provide the most direct insight, and form the basis of the discussion below. For example, Fig.~1 of Ref.~\cite{Albiez_etal05} and Fig.~1b of Ref.~\cite{Zimmermann_2018} provide clear depictions of the LMG model's spectrum in both phases.

In the symmetric phase the spectrum of $\Hh_S$ is nondegenerate, and therefore all eigenstates have well-defined parity by default. In the symmetry-broken phase the low-lying eigenstates occur in pairs with opposite parity and closely spaced eigenvalues, which become degenerate in the thermodynamic limit. This permits the construction of eigenstates that lack well-defined parity and have non-zero $S_x$ expectation values. At large but finite $S$ the ground and first excited states have even and odd parity respectively, and are separated by an energy gap $\Delta E$ which is exponentially small in $S$. From this quasi-degenerate pair it is possible to form linear combinations which are initially localized around one of the semi-classical (symmetry-broken) ground states. Under the unitary dynamics generated by $\Hh_S$ this leads to back-and-forth tunneling between these ground states, with a frequency of $\omega=\Delta E$. This scenario is familiar from the one-dimensional double-well potential. Here we can picture the dynamics as taking place on the Bloch sphere, with energy minima at the point(s) corresponding to the ground state magnetization $\bm{m}$ in \eqref{eq:magnetization}; see Fig.~1 of Ref.~\cite{Zibold2010} for an illustration.

The numerical results reported in Sec.~\ref{sec:numericsspectum} will demonstrate that essentially the same scenario plays out on the level of the Lindbladian. Here we summarize the main points. We define the action of the parity operator $P$ on $\rho$ by $\AdP \rho=P^\dag\rho P$, and the space $\Kk$ of state operators then splits into the direct sum of two subspaces $\Kk_+$ and $\Kk_-$. Elements of $\Kk_+$ obey $\AdP\rho=\rho$, and therefore preserve the parity of states they act on, while elements of $\Kk_-$ satisfy $\AdP\rho=-\rho$ and flip the parity of states. Since $P^\dag \Hh_S P=\Hh_S$, $P^\dag \Hh_\gamma P=\Hh_\gamma$ and $P^\dag L P=-L$ we see from \eqref{eq:lindblad1} that $\AdP$ commutes with the Lindbladian $\Ll$, and so each eigenoperator of $\Ll$ can be chosen to lie in either $\Kk_+$ or $\Kk_-$. In particular, the Lindblad evolution will not mix these two subspaces. In fact, if the stationary state is unique then it must be an element of $\Kk_+$, as the elements of $\Kk_-$ are traceless. For any $\rho\in\Kk_+$ we have $\Tr(S_x\rho)=0$, and such a stationary state therefore respects the symmetry present in $\Hh_S$. This agrees with what was found analytically for the symmetric phase of the LMG Hamiltonian.

In the symmetry-broken phase the quasi-degenerate pairing of odd and even parity $\Hh_S$ eigenstates results in a similar pairing of $\Ll$ eigenoperators from $\Kk_+$ and $\Kk_-$. Appropriate linear combinations of these pairs then produce state operators with support around one of the two semi-classical ground states. At finite $S$ this degeneracy is not exact and tunneling between these ground states still occurs, eventually leading back to the unique stationary state in $\Kk_+$. However, in the large-$S$ limit tunneling is completely suppressed, and the stationary state in $\Kk_+$ becomes degenerate with a state from $\Kk_-$. This allows for the construction of thermal states with support around one of the two semi-classical ground states. The results obtained by applying the bosonization procedure in the symmetry-broken phase therefore describe the fixed point of this local thermalization process, and are applicable either in the limit of large $S$ or, for finite $S$, on timescales far smaller than the tunneling time.

\section{Numerical Results}
\label{sec:numerics}

\begin{figure}
	\centering
	\includegraphics[width=0.9\linewidth]{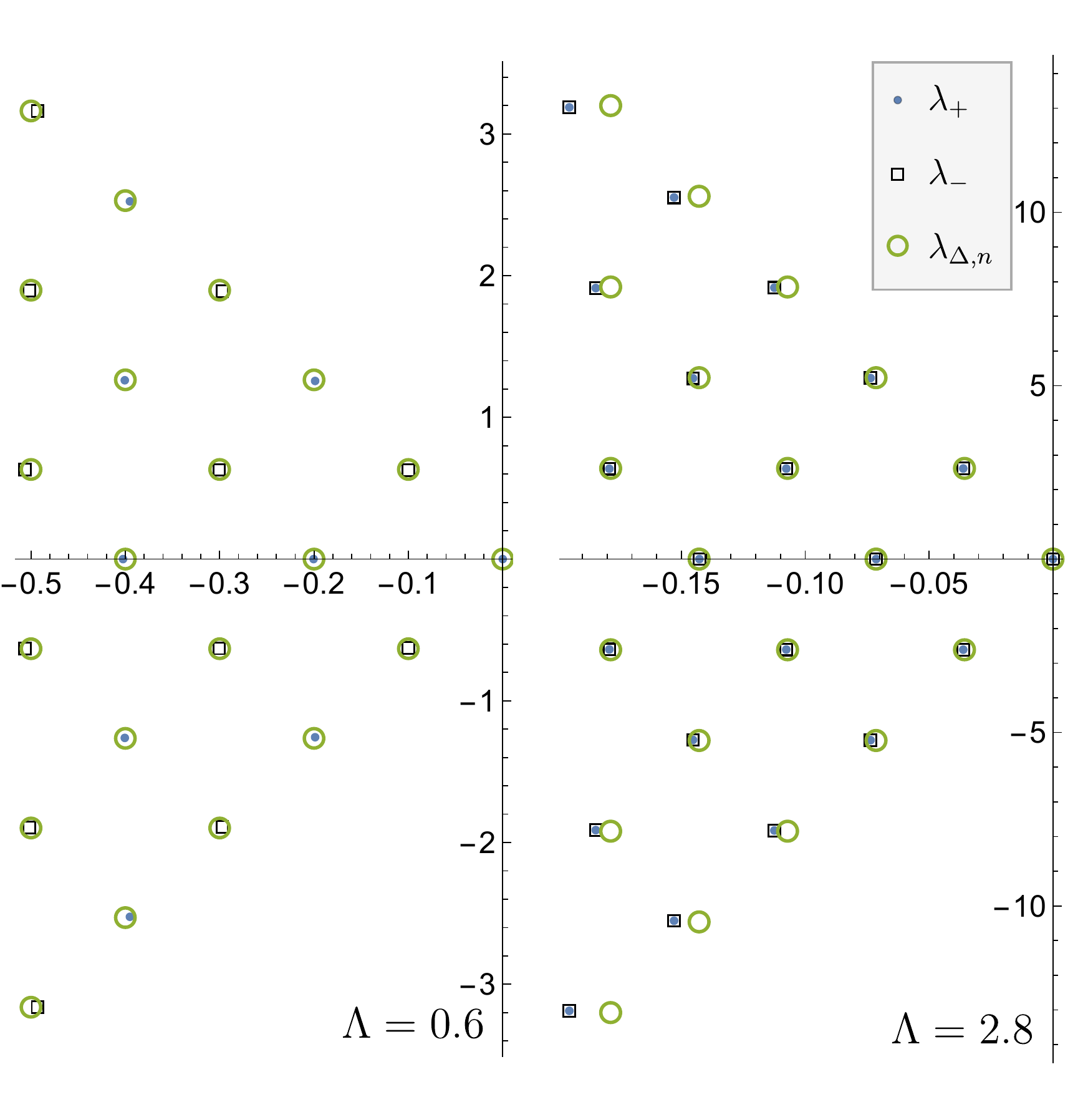}
	\caption{Spectrum of the Lindbladian \eqref{eq:lindblad1} in a region of the left complex plane, computed for parameter values $\gamma=0.2$ and $T=4$. The left and right subfigures show results for the symmetric ($\Lambda<1$) and symmetry-broken ($\Lambda>1$) phases respectively. Here $\lambda_\pm$ are numerical results for $S=3000$ for the eigenvalues corresponding to the $\Kk_\pm$ subspaces of $\Kk$, while $\lambda_{\Delta,n}$ are the predictions of equation \eqref{eq:Leigenvalues}, which is based on a Holstein-Primakoff approximation and assumes small coupling $\gamma$. \label{fig:spectrumplot}}
\end{figure}

The approximate analytic results of Sec.~\ref{s:bosonization} are obtained by truncating the Holstein-Primakoff (HP) transformation \eqref{eq:HPMapping} at suitable orders in the small parameter $1/S$, and the approximation is therefore valid only for sufficiently large spin quantum numbers $S$. A second restriction on the validity of the approximation is related to shape of the semi-classical potential that is approximated, which in turn is determined by the LMG Hamiltonian \eqref{eq:Hs}. The HP approximation replaces that original Hamiltonian by a harmonic oscillator Hamiltonian \eqref{eq:HSdiagonal}. The more the semi-classical potential associated with the original Hamiltonian differs from a parabola, the less accurate is the HP approximation. In the symmetry-broken phase, the semi-classical potential of the LMG model has a double-well structure and, while each of the wells separately can be approximated by a parabola, the overall shape of the potential can not, and the HP approximation is unable to capture any tunneling between the wells. The aim of the present section is to provide numerical results for the original spin Lindblad equation \eqref{eq:lindblad1} that allow us to assess the range of validity of the HP approximation in the various parameter regimes of the model.

\subsection{Spectrum of \texorpdfstring{$\Ll$}{L}}
\label{sec:numericsspectum}

The numerical data shown and discussed in this subsection mainly serve the purpose of justifying the claims about the properties of the eigenvalue spectrum of the Lindblad superoperator $\Ll$ made in Sec.~\ref{sec:symmetries}, in particular regarding the formation of near-degenerate pairs of eigenvalues. Figure \ref{fig:spectrumplot} shows a subset of the eigenvalues of the Lindbladian in a region of the left complex plane. The numerical calculations were performed by restricting the Hilbert space to the subspace spanned by the lowest $101$ eigenstates of $\Hh_S$. In the left panel of Fig.~\ref{fig:spectrumplot}, which shows data for the symmetric phase, we observe very good agreement of the numerical data based on the Lindbladian \eqref{eq:lindblad1} with the HP predictions of equation \eqref{eq:Leigenvalues}. In particular, there is a clear separation between the eigenvalues originating from $\Kk_+$ and $\Kk_-$. In the right panel of Fig.~\ref{fig:spectrumplot}, corresponding to the symmetry-broken phase, a rather different scenario is observed, with eigenvalues arranged into nearly degenerate pairs from $\Kk_+$ and $\Kk_-$. While the eigenvalues themselves still follow the trend predicted by Eq.~\eqref{eq:Leigenvalues}, it should be understood that the eigenstates given by \eqref{eq:Leigenstates} now correspond to particular linear combinations of these pairs.

\begin{figure}\centering
  \includegraphics[width=0.9\linewidth]{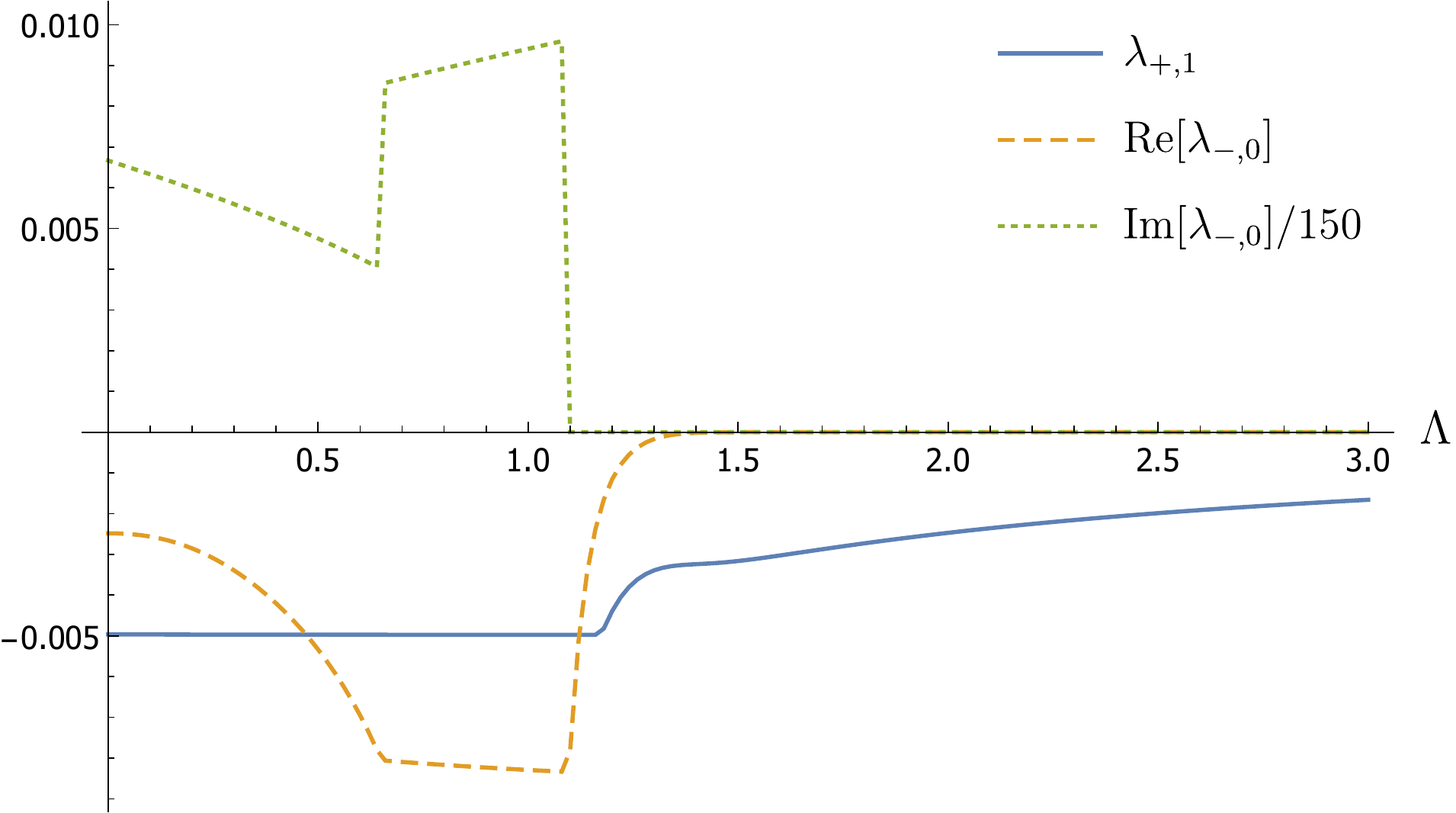}
	\caption{\label{fig:eigenvalueplot500}
The eigenvalue $\lambda_{+,1}$ of $\Ll$ with the largest non-zero real part in the $\Kk_+$ subspace of $\Kk$, shown as a function of $\Lambda$ and for parameter values $S=500$, $\gamma=0.005$, and $T=4$. Also shown are the real and imaginary parts of $\lambda_{-,0}$, the eigenstate from the $\Kk_-$ subspace with the largest real part. Note that $\lambda_{-,0}$ becomes nearly degenerate with $\lambda_{+,0}=0$ at large $\Lambda$.}
\end{figure}

\begin{figure}\centering
  \includegraphics[width=0.9\linewidth]{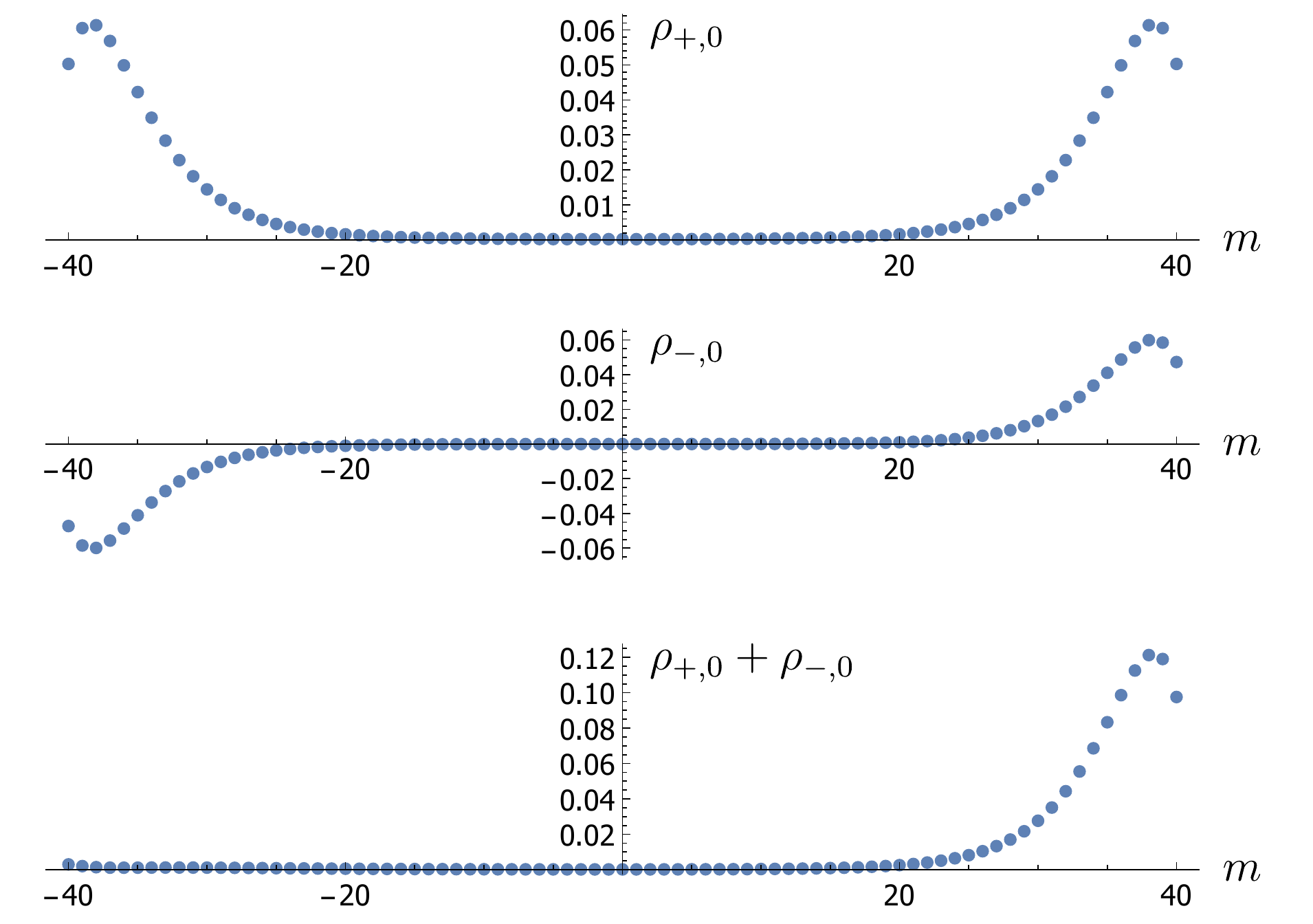}
	\caption{\label{fig:diagonalplot}
Diagonal entries of the stationary state $\rho_{+,0}$ (top) and nearly degenerate $\rho_{-,0}$ (center) eigenstate of $\Ll$ in the $S_x$ basis as functions of the eigenvalue of $S_x$ for $S=40$ and $\Lambda=2.6$. The sum $\rho_{+,0}+\rho_{-,0}$ (bottom) corresponds to a state operator with support around the classical ground state with positive $m_x$ magnetization.}
\end{figure}

For a quantitative analysis of the formation of eigenvalue pairs, we order the  eigenvalues from the $\Kk_+$ and $\Kk_-$ sectors such that $\myRe(\lambda_{\pm,0})>\myRe(\lambda_{\pm,1})>\myRe(\lambda_{\pm,2})>\cdots$, with $\lambda_{+,0}=0$ corresponding to the stationary state. (The eigenvalues of $\Ll$ occur in complex conjugate pairs. Here we disregard those with negative imaginary parts.) The equilibration rate at long times is set by the smaller of $|\myRe(\lambda_{+,1})|$ and $|\myRe(\lambda_{-,0})|$. Figure \ref{fig:eigenvalueplot500} shows $\lambda_{+,1}\in\mathbb{R}$ together with the real and imaginary parts of $\lambda_{-,0}$ as functions of $\Lambda$ \footnote{The numerical diagonalization is performed in a subspace of density operators which are eigenoperators of the unitary part of the Lindbladian, with eigenvalues within a certain range around zero. For smaller values of $S$ the results of this restricted diagonalization were bench-marked against exact diagonalization using the full state space.}. Note that due to finite size effects  $\lambda_{+,1}$ and $\lambda_{-,0}$ match the analytic predictions $\lambda_{0,1}$ and $\lambda_{1,0}$ of \eqref{eq:Leigenvalues} only for small $\Lambda$. However, here our goal is not to benchmark the analytic results, but to highlight the generic trends observed when crossing into the symmetry-broken phase. In particular, we see that there is a point, just beyond $\Lambda=1$, where $\lambda_{-,0}$ becomes real and rapidly approaches $\lambda_{+,0}=0$ with increasing $\Lambda$. This results in a very slow decay of the corresponding eigenoperator, and allows for the construction of a quasi-stationary state localized around one of the classical ground states.

Figure \ref{fig:diagonalplot} shows the diagonal entries of $\rho_{+,0}$ and $\rho_{-,0}$ in the $S_x$ basis, plotted as functions of the corresponding $S_x$ eigenvalue. With the parameter $\Lambda$ chosen well inside the symmetry-broken phase, we see the expected parity symmetry in $\rho_{+,0}$, with peaks around the two values of $m_x=\sin\theta_0$ associated with the semi-classical ground states. In contrast, $\rho_{-,0}$ is not a physical state operator, but it can be normalized so as to ensure that the combination $\rho_{+,0}\pm\rho_{-,0}$ is a physical state. The latter will have support around only one of the semi-classical ground states. It is this locally thermalized, symmetry-broken state that the bosonized large-$S$ calculations yield as a stationary state in \eqref{eq:rho00vector}.

\subsection{Dynamics}
\label{sec:dynamics}

The near-degenerate eigenvalue pairs discussed in Sec.~\ref{sec:numericsspectum}, and the resulting double-peak structure illustrated in Fig.~\ref{fig:diagonalplot}, give rise to tunneling dynamics between negative-$m$ and positive-$m$ states, corresponding to the two wells of the semi classical potential. As mentioned at the beginning of Sec.~\ref{sec:numerics}, this tunneling dynamics is not captured by the Hol\-stein-Pri\-ma\-koff approximation of Sec.~\ref{s:bosonization}, and the present section is devoted to numerically analyzing the tunneling on the basis of the original Lindblad equation \eqref{eq:lindblad1}, which in turn will provide insights into the time scales after which the bosonization results of Sec.~\ref{s:bosonization} are bound to fail.

\begin{figure}\centering
  \includegraphics[width=0.9\linewidth]{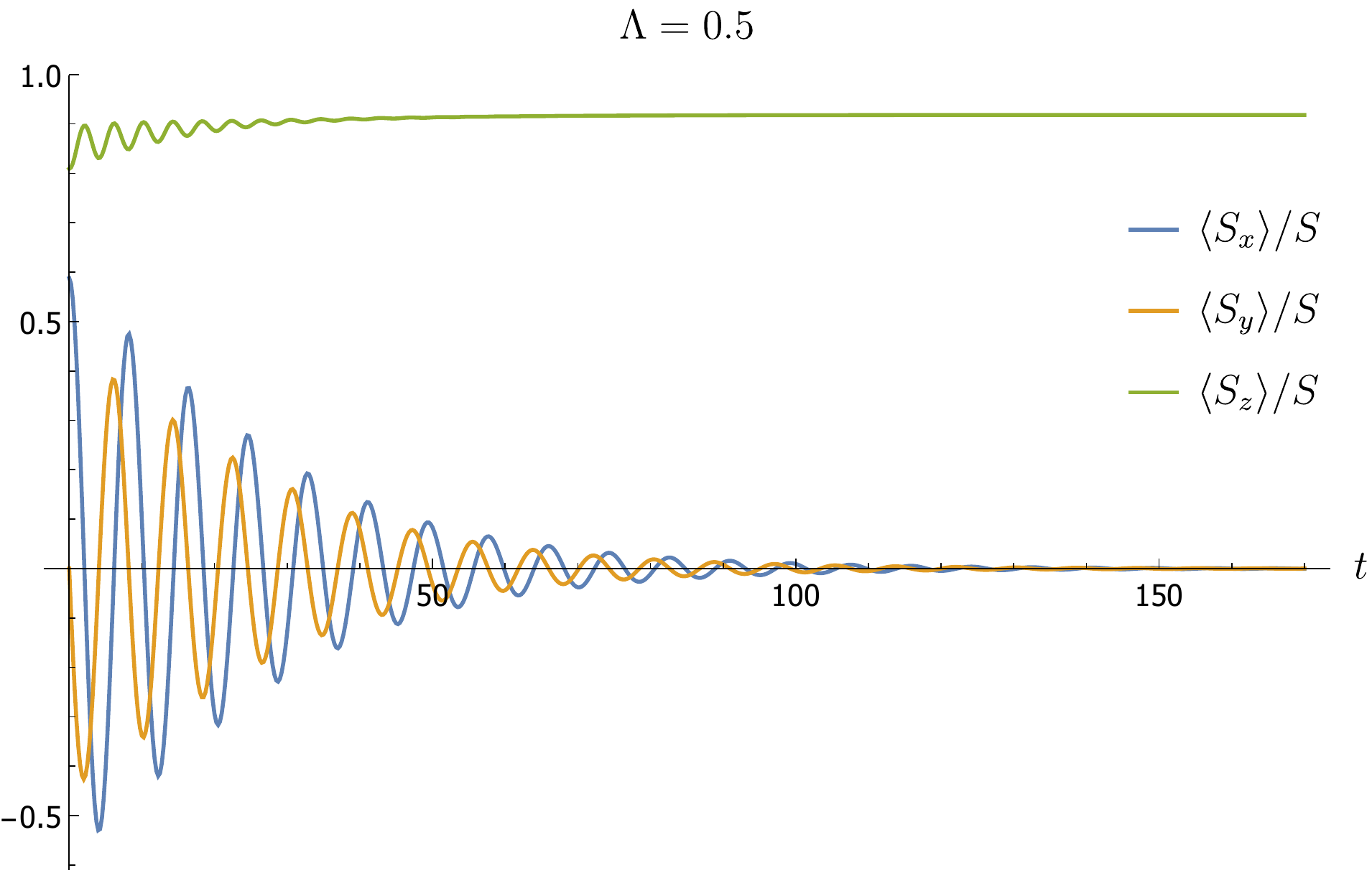}
  \includegraphics[width=0.9\linewidth]{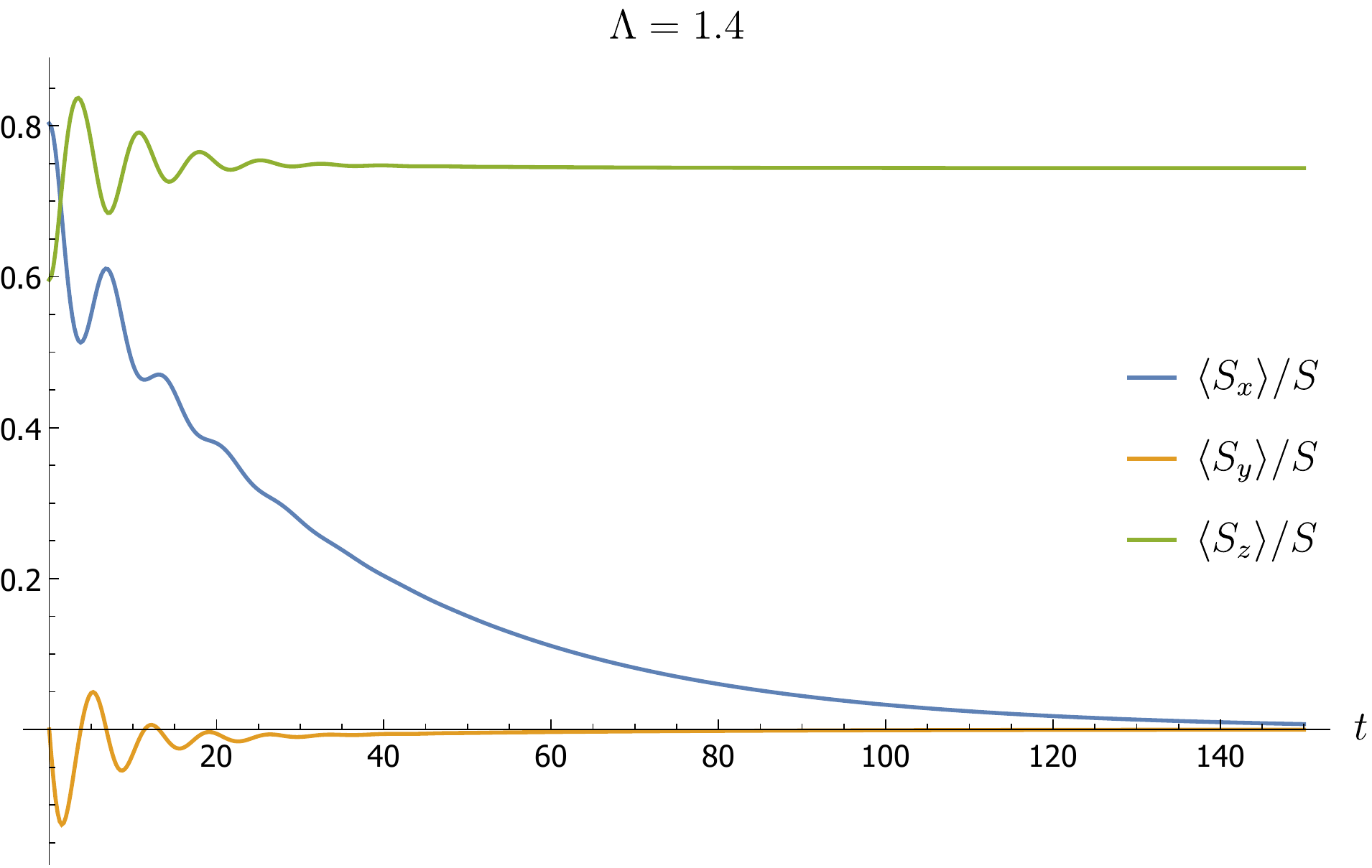}
	\includegraphics[width=0.9\linewidth]{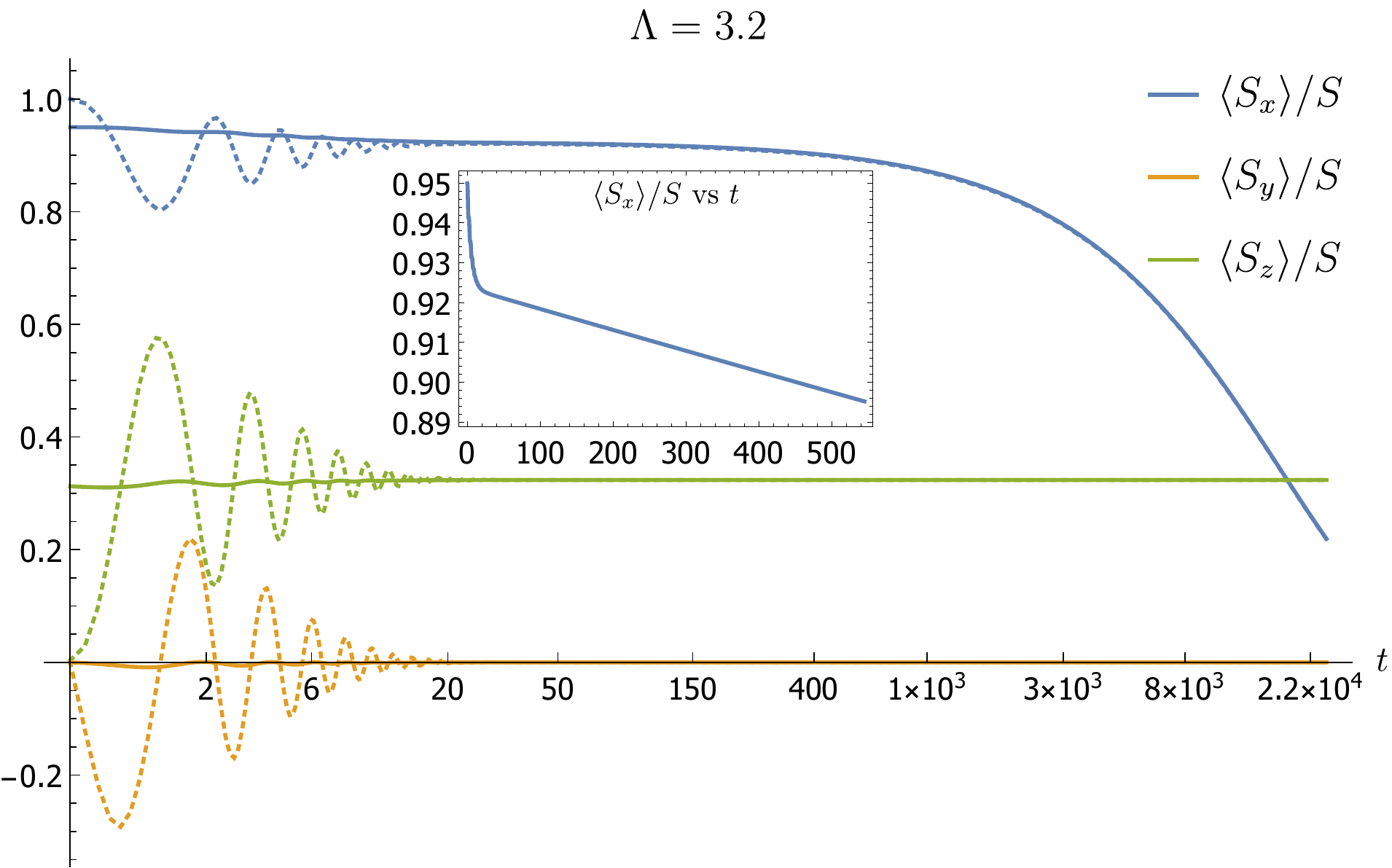}
	\caption{\label{f:dynamics}
Exact numeric results for the dynamics of $\ex{\bm{S}}$, obtained by solving the original Lindblad equation \eqref{eq:lindblad1} with $S=60$. The choices of initial states are described in the text. Top: Parameter values $\gamma=0.05$ and $T=4$, with $\Lambda=0.5$ in the symmetric phase. Center: As in the top panel, but for $\Lambda=1.4$, which is slightly inside the symmetry-broken phase. Bottom: For parameter values $\gamma=0.64$ and $T=5$, with $\Lambda=3.2$, a point deep inside the symmetry-broken phase. The dashed and solid lines correspond to different choices of initial state, as explained in the text.}
\end{figure}

Our choice of initial state is a simplified version of that considered in Eqs.~\eqref{e:rho0prime}--\eqref{e:Ry} of section \ref{sec:timeevolutionofdensitymatrix}. Rather than considering the rotated ground state of $\Hh_S$, we set
\begin{equation}\label{e:rho0}
\rho(0)=\ket{\psi}\bra{\psi}
\end{equation}
with
\begin{equation}
\ket{\psi}=R_y(\theta)\ket{S,S},
\end{equation}
where $\ket{S,S}$ is the $S_z$ eigenstate satisfying
\begin{equation}\label{e:SS}
S_z\ket{S,S}=S\ket{S,S}.
\end{equation}

The initial spin orientation is therefore in the $\bm{m}(0)=(\cos\theta,0,\sin\theta)$ direction. Choosing $\theta=\theta_0$ as in Eq.~\eqref{e:theta0} will align $\bm{m}(0)$ with the ground state magnetization \eqref{eq:magnetization}, which amounts to localizing the initial state in the corresponding minimum of the semi-classical potential. In the following we study the dynamics and the pertinent timescales for three exemplary cases, corresponding to the symmetric phase, the weakly symmetry-broken phase, and the strongly symmetry-broken phase.

Figure \ref{f:dynamics} (top) shows the dynamics for a parameter value $\Lambda=0.5$ in the symmetric phase and for an initial state \eqref{e:rho0}--\eqref{e:SS} rotated out of the $z$-direction by $\theta=\pi/5$. The time dependence of the components of $\ex{\bm{S}}=\Tr(\rho\bm{S})$ shows evolution on two distinct timescales, namely a slow relaxation to the corresponding equilibrium values, superimposed by fast oscillatory behavior. The initial misalignment between $\bm{m}_0$ and the equilibrium magnetization $\bm{m}=(0,0,1)$ results in oscillations with a frequency close to $\omega_b\approx 0.7$, decaying at a rate set by $\myRe(\lambda_{-,0})\approx -0.042$.

Dynamics in the symmetry-broken phase is shown in Fig.~\ref{f:dynamics} (center), for which we chose $\Lambda=1.4$ and an initial spin orientation rotated by an angle $\theta-\theta_0=\pi/20$ away from the positive $m_x$ semi-classical ground state magnetization. Qualitatively $\ex{S_y}$ and $\ex{S_z}$ behave similar to the top panel of Fig.~\ref{f:dynamics}, with the difference that $\ex{S_x}$ approaches its equilibrium value $0$ on a much longer timescale. The reason for this slow decay is that tunneling between the two classical ground states is required for the system to equilibrate, and the timescale associated with this tunneling is set by the inverse gap of the Lindbladian spectrum, $1/\myRe(\lambda_{-,0})\approx 33$. For the parameter values used in here, the {\em Hamiltonian}\/ energy gap $\Delta E$ between the ground state and the first excited state of $\Hh_S$ is several orders of magnitude smaller than the Lindbladian gap $\myRe(\lambda_{-,0})$. This implies that tunneling due to the unitary dynamics generated by $\Hh_S$ occurs on a significantly longer timescale than what is observed here, and that the decay of $\ex{S_x}$ is therefore dominated by the dissipative part $\Dd$ \eqref{e:D} in the Lindbladian. 

Dynamics further inside the symmetry-broken phase at $\Lambda=3.2$ is shown in Fig.~\ref{f:dynamics} (bottom). The solid lines correspond to data for an initial spin orientation aligned with $\bm{m}$, the dashed lines are for an initial deviation from $\bm{m}$ by an angle $\theta-\theta_0=\pi/10$. As expected, the former case leads to less pronounced oscillations than the latter. In both cases, all three spin components appear to approach constant values which are independent of the specific initial state. This is an illustration of the local thermalization process occurring around one of the symmetry-broken semi-classical ground states, and this local thermalization is well described by the results of Secs.~\ref{sec:lindbladdiagonalise} and \ref{sec:stationarystate}. While the $S_y$ and $S_z$ components have indeed reached their equilibrium values, the $S_x$ component is still undergoing a very slow exponential decay to zero, which is only apparent over long time scales. Note the non-linear scale on the horizontal axis. The inset on the bottom plot in Fig.~\ref{f:dynamics} shows $\ex{S_x}$ over a shorter time interval. The change in slope seen at around $t\approx 50$ indicate the cross-over from fast local thermalization to the slow approach to the true thermal stationary state, a process that cannot be captured by the bosonization methods of Sec.~\ref{s:bosonization}.

\section{Conclusions}
On the methodological side, the main result of the present paper is an approximate analytical method, based on bo\-son\-i\-za\-tion and vectorization techniques. The method is presented for the case of a Markovian dissipative Lipkin-Meshkov-Glick model, but is more generally applicable to Lindbladians of large-S spin models.
This method approximately maps the spin Lindblad master equation \eqref{e:unitary}--\eqref{eq:lindblad1} onto a bosonic Lindblad master equation defined by \eqref{eq:lindblad1} with \eqref{eq:Hsa}, \eqref{eq:finalHgamma}, and \eqref{eq:finalL}. This equation, which is quadratic in the bo\-son\-ic operators, can then be tackled either by exact or by approximate asymptotic methods. An exact solution of the quadratic Lind\-blad\-ian is reported in Sec.~\ref{sec:strongcoupling} for arbitrary dissipation strength $\gamma$ by employing the method of third quantization. A simpler, more manageable closed-form solution obtained by perturbation theory in the weak-dissipation limit is reported in Sec.~\ref{sec:lindbladdiagonalise}.

The simplicity of these results relies on the approximation made when truncating the Taylor series expansion of the bosonization (Holstein-Primakoff) mapping \eqref{eq:HPMapping} at leading order in $1/S$. The validity and accuracy of the method therefore depends firstly on the spin quantum number $S$ being large, but also, more subtly, on the range of validity of the Taylor expansion, which is crucially affected by whether or not the underlying semi-classical potential of the system Hamiltonian $\Hh_S$ is well approximated by a parabola. In a dynamical context, it furthermore becomes relevant whether the system's initial state lies within the range of validity of the quadratic approximation, and whether the state evolves within that range at later times. While this is in general a difficult question to answer, the numerical results of Sec.~\ref{sec:numerics} provide at least guidelines for assessing that region of validity. Compared to other large-$S$ analytic techniques for the dissipative LMG model, like those put forward in Ref.~\cite{LouwKrielKastner19}, the methods developed in the present paper have the desirable feature of including leading-order finite-size corrections. This not only leads to a more accurate results for large, finite systems as they are potentially relevant for experimental realizations of the LMG model, but also circumvents some of the pathologies of the strict infinite-$S$ system that were discussed in Ref.~\cite{Webster2018}.

Beyond method development, our work provides insights into the physics of thermalization in open quantum systems. A quantum system coupled to a bath of temperature $T$ is in general not guaranteed to evolve towards a Gibbs canonical equilibrium state \cite{PopescuShortWinter06,Subasi_etal12}. When deriving a Markovian master equation describing such a system, in many cases a secular approximation is performed, which has the merit of guaranteeing complete positivity of the quantum dynamical semigroup, but also essentially enforces a Gibbs canonical equilibrium state as the stationary state of the master equation. To retain the possibility of more diverse equilibrium properties, we investigated the master equation specified in Eqs.~\eqref{e:Lindblad_general}--\eqref{eq:lindblad1} that has been derived without a secular approximation, but on which complete positivity has been enforced by an alternative method (see Appendices A and B of Ref.~\cite{LouwKrielKastner19} for details). The study of the equilibrium state and the equilibration dynamics of that master equation is therefore a nontrivial problem that may contribute to the understanding of equilibration in open quantum systems. By applying the bosonization and vectorization techniques developed in Sec.~\ref{s:bosonization}, we found that the equilibrium state \eqref{eq:rho00} of the dissipative LMG model has the functional form of a Gibbs thermal state proportional to $\exp(-\Hh_S/T_\text{ss})$, but with a stationary temperature $T_\text{ss}$ \eqref{e:Tss} that in general differs from the ``imposed'' bath temperature $T$. When studying the dynamical approach of the equilibrium state we observed that the time-evolved density operator $\rho(t)$ \eqref{eq:evolvedrho}--\eqref{eq:TSdefinition} equilibrates by passing through a continuum of thermal states on which damped oscillations are superimposed. This is reminiscent of quasi-adiabatic evolution, but differs from conventional adiabatic dynamics in that the time evolution is not driven by a slowly varying parameter.

Extensions of the present work should aim to address the restrictive nature of the initial Holstein-Primakoff mapping from the spin to bosonic degrees of freedom. The essentially local nature of this approximation rules out any description of the tunneling effects which are integral to the equilibration process in the symmetry broken phase. It would be interesting to seek generalizations of this mapping, possibly involving more than one species of boson, which could capture the non-local dynamics resulting from the double-well shape of the classical potential energy. If such a mapping resulted in a quadratic bosonised Lindblad equation then the methods presented here, and that of the third quantization approach \cite{prosen2010}, would provide a versatile toolkit for further analysis.

\appendix*

\section{\texorpdfstring{$\mathrm{su}(1,1)$}{su(1,1)} Transformations}
Some of the calculations in the Holstein-Primakoff boson\-ized version of the dissipative LMG model are most conveniently performed by exploiting the two-mode representation of an underlying $\mathrm{su}(1,1)$ algebraic structure. The details of these calculations are reported in the following appendices.

\subsection{Diagonalization}
\label{sec:su11diagonalisation}
Here we outline the construction of the similarity transformation $\Tt$ used in Sec.~\ref{sec:lindbladdiagonalise} to bring the projected perturbative term
\begin{multline}
\Ll'_\Delta=m_z/2-(B_+^2+B_-^2)\frac{1}{2}[b_1^\dag b_1+b_2^\dag b_2+1]
\\+B_+^2b_1^\dag b_2^\dag+B_-^2 b_1 b_2
\end{multline}
into a diagonal form. Note that $\Ll'_\Delta$ can be expressed as 
\begin{equation}
\Ll'_\Delta=m_z/2-(B_+^2+B_-^2)K_0+B_+^2K_++B_-^2 K_-,
\end{equation}
where
\begin{subequations}
\begin{align}
	K_0&=\frac{1}{2}[b_1^\dag b_1+b_2^\dag b_2+1],\\
	K_-&=b_1 b_2,\\
	K_+&=b_1^\dag b_2^\dag.
\end{align}
\end{subequations}
These operators obey the $\mathrm{su}(1,1)$ commutation relations 
\begin{align}
	[K_0,K_\pm]&=\pm K_\pm,& [K_-,K_+]&=2K_0.
\end{align}
We construct the desired transformation $\Tt$ in two steps. First we apply the similarity transformation 
\begin{equation}
	\Tt_1=e^{-\ln(B_-/B_+)K_0},
	\label{eq:T1def}
\end{equation}
which leaves $K_0$ unchanged but rescales $K_\pm$ as
\begin{equation}
	\Tt_1^{-1}K_\pm \Tt_1=\left(\frac{B_-}{B_+}\right)^{\pm 1}K_\pm.\label{eq:T1Identity}
\end{equation}
This, and subsequent, identities can be verified using Baker-Campbell-Hausdorff-type formulas, or more easily by using the $2\times 2$ representation of $\mathrm{su}(1,1)$ in terms of Pauli matrices as $K_\pm=\mp\sigma_\pm$ and $K_0=\frac{1}{2}\sigma_z$. This check is sufficient, since identities which rely only on the algebraic properties of these operators can be verified using any faithful representation. Applying $\Tt_1$ to $\Ll'_\Delta$ gives the Hermitian pairing problem
\begin{equation}
	\Tt_1^{-1}\Ll'_\Delta \Tt_1=m_z/2-(B_+^2+B_-^2)K_0+B_+B_-(K_++K_-).
\end{equation}
This operator can be diagonalized by the unitary transformation
\begin{equation}
	\Tt_2=\exp(i\psi K_2),
	\label{eq:T2def}
\end{equation}
where $K_2=-i(K_+-K_-)/2$ and 
\begin{equation}
	\tanh\psi=\frac{2B_+B_-}{B_+^2+B_-^2}.
\end{equation}
This leads to the final form
\begin{equation}
	\Tt_2^{-1}\Tt_1^{-1}\Ll'_\Delta \Tt_1 \Tt_2=-\frac{m_z}{2}\left(b_1^\dag b_1+b_2^\dag b_2\right),
\end{equation}
where \eqref{eq:BIdentity} was used. Finally, we combine the two transformations into $\Tt=\Tt_1\Tt_2$.

\subsection{Factorization}
\label{sec:su11factorisation}
Here we summarize the steps which produce the simplified form of the vectorized stationary state $\Tt\ket{0,0}=\Tt_1\Tt_2\ket{0,0}$ given in Eq.~\eqref{eq:rho00vector}. The transformation \eqref{eq:T2def} can be factorized as \cite{Santiago_1976}
\begin{equation}
	\Tt_2=e^{\tanh(\psi/2) K_+}e^{\ln\left[\text{sech}^2(\psi/2)\right]K_0}e^{-\tanh(\psi/2)K_-},
\end{equation}
where
\begin{align}
	\tanh\left(\frac{\psi}{2}\right)&=\frac{B_+}{B_-},&\text{sech}^2\left(\frac{\psi}{2}\right)&=1-\frac{B_+^2}{B_-^2}.
\end{align}
Since $K_-\ket{0,0}=0$ and $K_0\ket{0,0}=\frac{1}{2}\ket{0,0}$, we read off that
\begin{equation}
	\ket{\rho_{0,0}}=\Tt_1\Tt_2\ket{0,0}\propto\Tt_1 e^{(B_+/B_-)K_+}\ket{0,0}.
\end{equation}
Using \eqref{eq:T1Identity} leads to the final form
\begin{equation}
	\ket{\rho_{0,0}}\propto e^{(B_+/B_-)^2K_+}\ket{0,0}.
\end{equation}

\subsection{Evolving the state operator}
\label{sec:su11evolution}
In this section we report the calculation of the time-evolved density operator given by Eqs.~\eqref{eq:evolvedrho}--\eqref{eq:TSdefinition}. The pure initial state is characterized by the vector $\ket{\psi}$ in Eq.~\eqref{eq:psi}, which, upon application of the HP mapping \eqref{eq:HPMapping} to lowest order in $1/S$ and the Bogoliubov transformation in \eqref{eq:bogoliubov}, can be written as
\begin{equation}
	\ket{\psi}=e^{\theta \sqrt{S/2} (a^\dag-a)}\ket{0}_b=e^{\theta' (b^\dag-b)}\ket{0}_b
\end{equation}
with $\ket{0}_b$ the $b$-boson vacuum and $\theta'=\theta \sqrt{S/2} \exp(-\phi_b/2)$. Upon vectorizing $\rho(0)$ we obtain
\begin{equation}
	\ket{\rho(0)}=e^{\theta'\left(b_1^\dag-b_2+b^\dag_2-b_1\right)}\ket{0,0}_b.
\end{equation}
To evolve $\ket{\rho(0)}$ in time we apply the operator $\exp(t\Ll_\Delta)=\exp(t\Ll_0)\exp(t\gamma\Ll'_\Delta)$ with $\Ll_0$ and $\Ll'_\Delta$ given in Eqs.~\eqref{eq:L0Def} and \eqref{eq:LPrimeDef}. We do so in three steps. First we apply $\exp(t\gamma\Ll'_\Delta)$ and use the fact that
\begin{equation}
	e^{t\gamma\Ll'_\Delta}\bigl(b_i^\dag-b_j\bigr)e^{-t\gamma\Ll'_\Delta}=e^{-m_z\gamma t/2}\bigl(b_i^\dag-b_j\bigr)
\end{equation}
where $(i,j)$ is $(1,2)$ or $(2,1)$. This produces
\begin{equation}
	e^{t\gamma\Ll'_\Delta}\ket{\rho(0)}=e^{\theta''\left(b_1^\dag-b_2+b^\dag_2-b_1\right)}e^{t\gamma\Ll'_\Delta}\ket{0,0}_b,
	\label{eq:evolvestep1}
\end{equation}
where $\theta''=\theta'\exp(-m_z\gamma t/2)$. Next we simplify $\exp(t\gamma\Ll'_\Delta)\ket{0,0}_b$ by using the factorization
\begin{equation}
	e^{t\gamma\Ll'_\Delta}=e^{\gamma m_z t/2}e^{A_+ K_+} e^{2\ln(A_0) K_0}e^{A_- K_-}
\end{equation}
with
\begin{align}
	A_+&=B_+\left(B_-^2+\frac{m_z}{e^{m_z\gamma t}-1}\right)^{-1},\\
	A_0&=e^{-m_z\gamma t/2}(1-A_+),
\end{align}
where Eq.~\eqref{eq:BIdentity} has been used. This allows us to write
\begin{equation}
	e^{t\gamma\Ll'_\Delta}\ket{0,0}_b=(1-A_+)e^{A_+ K_+}\ket{0,0}_b,
\end{equation}
which is then inserted into \eqref{eq:evolvestep1}. What remains is to apply $\exp(t\Ll_0)$ to the expression in \eqref{eq:evolvestep1}. Since $\Ll_0$ contains only $b$-boson number operators, this is straightforward. The final expression for $\ket{\rho(t)}=\exp(t\Ll_\Delta)\ket{\rho(0)}$ is
\begin{multline}\label{e:rhotfinal}
	\ket{\rho(t)}=(1-A_+)\exp\left\{\theta''\left[e^{-i\omega_b t}(b_1^\dag-b_2)-\text{h.c.}\right]\right\}\\
	\times \exp\left(A_+ K_+\right)\ket{0,0}_b.
\end{multline}
Following the same steps as in Sec.~\ref{sec:stationarystate} to return from the vectorized form to the operator description yields $\rho(t)$ as given in Eq.~\eqref{eq:evolvedrho}.

\bibliography{ThermRef}

\end{document}